\documentclass[11pt]{article}

\usepackage[T1]{fontenc}
\usepackage[utf8]{inputenc}
\usepackage{microtype}

\usepackage[width=150mm,top=28mm,bottom=25mm]{geometry}

\usepackage{graphicx}
\graphicspath{{images/}}
\usepackage{caption}
\usepackage{subcaption}
\usepackage{float}

\usepackage{booktabs,longtable}
\usepackage[table]{xcolor}
\usepackage{colortbl}

\usepackage{amsmath,amssymb,amsfonts,amsthm}
\usepackage{mathtools}
\usepackage{extpfeil}

\DeclareMathOperator{\Var}{Var}
\DeclareMathOperator{\AVAR}{AVAR}

\DeclareMathOperator{\Corr}{Corr}

\theoremstyle{plain}

\theoremstyle{definition}

\theoremstyle{remark}

\newtheorem*{remark*}{Remark}

\newcommand{\para}[1]{\medskip\noindent\textbf{#1}\ }

\newcommand{\Ftrue}[1]{F_{[#1]}}
\newcommand{\Strue}[1]{S_{[#1]}}

\newcommand{\RE}{\mathrm{RE}}

\usepackage{csquotes}
\usepackage[
backend=biber,
style=authoryear,
sorting=nyt,
natbib=true,
dashed=false  
]{biblatex}
\usepackage{hyperref}
\usepackage{setspace}
\usepackage{fancyhdr}
\newif\ifanonymous
\anonymousfalse

\newcommand{\paperauthors}{Nabil Awan and Richard J.\ Chappell}
\newcommand{\paperaffil}{Department of Statistics, University of Wisconsin--Madison, WI, USA}

\title{Survival Analysis under Ranked Set Sampling}

\ifanonymous
\author{}
\else
\author{\paperauthors\\[4pt]\paperaffil}
\fi

\date{12 November 2025}

\begin{document}
	
	{\centering
		\textbf{Survival Analysis under Ranked Set Sampling}\\[6pt]
		\ifanonymous
		\else
		\paperauthors\\[4pt]
		\paperaffil\\[4pt]
		\fi
		\par
	}
	
	\vspace{1em}
	
\section{Specific Aims}

Ranked set sampling (RSS) is a cost-efficient design that uses inexpensive baseline ranking to select a more informative subset of individuals for full measurement \citep{McIntyre1952,Dell1972,Chen2004}. For uncensored outcomes, RSS is well known to deliver estimators with smaller variance than simple random sampling (SRS) of the same size, even when rankings are only moderately accurate. In survival analysis, however, existing RSS work has largely stopped at \emph{estimation} of the Kaplan--Meier (KM) curve under random censoring \citep{StrzalkowskaKominiak2014,mahdizadeh2017resampling}. Routine tools that applied researchers rely on for SRS data --- log-rank and weighted log-rank tests, restricted mean survival time (RMST/RML), and window mean life (WML) summaries --- are not yet available in a ranked-set setting, especially when ranking is imperfect and censoring is present.

The overarching goal of this thesis is to build a unified, RSS-based survival toolkit that preserves the efficiency gains of RSS while providing the same kinds of inferential outputs (tests, confidence intervals, interpretable functionals) that practitioners expect from SRS-based survival analysis.

\textbf{Aim 1 (estimation under RSS).} Formalize Kaplan--Meier and Nelson--Aalen estimators for right-censored data collected under balanced RSS, under both perfect and imperfect (concomitant-based) ranking. Establish their large-sample properties by adapting the random-censorship martingale/empirical-process framework of \citet{Stute1993,Stute1995} to the rank-wise RSS layout of \citet{StrzalkowskaKominiak2014}.

\textbf{Aim 2 (variance and efficiency).} Develop rank-aware, Greenwood-type variance estimators for RSS survival curves, and quantify efficiency gains relative to SRS through simulation grids that vary the set size $k$, number of cycles $m$, censoring proportion, and strength of the concomitant (e.g. Dell--Clutter-style misranking). Show when RSS meaningfully improves precision and when it collapses back to SRS.

\textbf{Aim 3 (extensions for practice).} Extend the RSS survival framework to standard tools that are currently missing: (i) log-rank and Fleming--Harrington-type weighted log-rank tests built from RSS risk and event processes; (ii) RSS versions of restricted mean life (RML) and window mean life (WML), with asymptotic variance formulas and two-sample comparisons; and (iii) an implementation plan (real-data illustrations and an \textsf{R} package) so the methods can be used in oncology, epidemiology, and reliability settings.

Collectively, these aims will move RSS for survival from “we can estimate the curve” to “we can analyze, compare, and report survival the way practitioners already do, but more efficiently.”

\section{Background and Significance}

\para{Ranked set sampling.}
RSS was introduced by \citet{McIntyre1952} in agricultural yield studies, based on a simple observation: ranking small sets of units is often cheap, but fully measuring all of them is expensive. In a balanced RSS design, we repeatedly draw sets of size $k$, rank the $k$ units using judgment or a cheap surrogate, and in the $r$-th draw of the cycle we \emph{only measure} the unit judged to have rank $r$, $r=1,\dots,k$. Repeating this for $m$ cycles yields $n = mk$ fully observed units,
\[
X_{[r]j}, \quad r=1,\dots,k,\quad j=1,\dots,m,
\]
where $X_{[r]j}$ denotes the unit judged to be the $r$-th order statistic in cycle $j$. Under \emph{perfect} ranking, this order coincides with the true order in each set; under \emph{imperfect} ranking (the practically relevant case), ranking is performed on a noisy concomitant $\tilde X = X + Z$ so the judged order only correlates with the truth.

The key fact, made precise in \citet{Dell1972}, is that RSS forces the measured sample to be spread across the entire support of the underlying distribution. As a result, the RSS sample mean is unbiased and typically has substantially smaller variance than the SRS mean of the same size, sometimes by 30--50\% or more depending on the distribution and $k$. Later work, summarized in \citet{Chen2004}, showed that this efficiency gain is \emph{robust}: even when rankings are not perfect but “better than random”, RSS still outperforms SRS.

\para{Why RSS is attractive for survival.}
In time-to-event studies, it is common to have a relatively cheap way to order subjects by their likely risk (baseline biomarkers, imaging, clinical scores, wearables) and a more expensive or longer process to obtain the actual event or censoring time. This creates exactly the same asymmetry that motivated RSS in agriculture: ranking is cheap, full follow-up is costly. An RSS design lets investigators (i) form small sets of eligible subjects, (ii) rank them by expected time-to-event using the auxiliary information, and (iii) fully follow only the selected unit from each set. This “rank-then-follow” workflow is realistic in oncology (imaging or marker-based ranking, then follow-up for recurrence), cardiovascular cohorts (lipids, blood pressure, ECG features → follow-up for MI or stroke), infectious-disease surveillance (symptom profile → follow-up for time-to-confirmed diagnosis), even business analytics (engagement → time-to-churn) --- all domains in which events can be rare, delayed, or administratively censored.

Compared to classical stratified sampling, RSS has an important advantage in these settings. Stratification over a continuous auxiliary variable requires choosing cut-points; bad cut-points waste information. RSS instead uses \emph{local} ranking within sets, preserving the full ordinal information of the auxiliary measure and avoiding a global partition. Under perfect ranking, this can deliver variance reductions approaching $1/k$; under imperfect ranking, the gain decays smoothly with the surrogate’s correlation, as documented in \citet{StrzalkowskaKominiak2014} and \citet{mahdizadeh2017resampling} for censored data. Thus, if a study can afford to form sets before committing to full follow-up, RSS is often strictly more informative than SRS for the same number of fully followed subjects.

\para{RSS with censored survival.}
For right-censored data, the survival target is
\[
S(t) = P(X > t),
\]
but we only observe $Y = \min(X,C)$ and $\delta = I(X \le C)$. Under SRS and independent censoring, the Kaplan--Meier estimator
\[
\widehat S(t) = \prod_{u \le t} \left(1 - \frac{dN(u)}{R(u)} \right)
\]
is the workhorse, and its properties are well understood: almost sure uniform convergence \citep{Stute1993}, central limit theorems for a wide class of functionals \citep{Stute1995}, and martingale representations \citep{FlemingHarrington1991,AndersenBorganGillKeiding1993}.

\citet{StrzalkowskaKominiak2014} were the first to carry this over to RSS with random censoring. In their setup, we still have a balanced RSS design with $k$ ranks and $m$ cycles, so the observed data are
\[
(Y_{[r]j}, \delta_{[r]j}), \quad r=1,\dots,k,\quad j=1,\dots,m,
\]
and within each rank $r$ these pairs are i.i.d. but come from a \emph{rank-specific} distribution (because higher-ranked units tend to have larger underlying $X$). Their RSS Kaplan--Meier estimator pools the rank-wise risk sets:
\[
\widehat S_{\text{RSS}}(t)
= \prod_{u \le t} \left( 1 - \frac{\sum_{r=1}^k dN_{[r]}(u)}{\sum_{r=1}^k R_{[r]}(u)} \right),
\]
which is exactly the SRS product-limit form but with rank-indexed counting and at-risk processes. They showed that this estimator is unbiased at event times, uniformly consistent, and asymptotically normal, and that in simulations it can have noticeably smaller variance than the SRS KM for the same $n = mk$. \citet{mahdizadeh2017resampling} then relaxed the perfect-ranking assumption by letting judgment be based on a noisy concomitant; using bootstrap-type procedures they showed that efficiency degrades \emph{smoothly} as ranking worsens, rather than collapsing immediately.

There are also side branches of this literature: parametric RSS survival with Type I censoring \citep{Yu2002}, PROS-based KM for incomplete rankings \citep{Nematolahi2020}, and several mean residual life (MRL) papers under (generally) fully observed RSS lifetimes \citep{ZamanzadeParvardehAsadi2019,ZamanzadeMahdizadehSamawi2024,ZamanzadeZamanzadeParvardeh2024}. These confirm the same message: whenever you can rank cheaply, RSS buys efficiency.

\para{What is still missing.}
Despite this progress, two gaps remain clear. First, almost all RSS-with-censoring papers stop at \emph{estimation} of $S(t)$ and all RSS-without-censoring papers at MRL-type functionals; they do not develop RSS analogues of the standard survival \emph{tests}, especially the log-rank and Fleming--Harrington family that dominate applied work. Second, the functionals that are most useful when proportional hazards fails --- restricted mean life (RML/RMST) and window mean life (WML), as in \citet{Paukner2021,Paukner2022,Paukner2023} --- have not yet been worked out under censored RSS, even though the RSS KM already exists. This thesis is motivated precisely by these two gaps.

\section{Innovation}

This proposal is innovative in three intertwined ways.

\para{1. RSS-aligned survival theory.}
We recast the RSS KM / NA estimators in the same empirical-process and martingale framework used by \textcite{Stute1993,Stute1995,AndersenBorganGillKeiding1993}, with the rank index carried through the influence-function representation. This provides a common language in which SRS, perfectly ranked RSS, and judgment-ranked RSS can be compared on $[0,\tau-\varepsilon]$.

\para{2. Rank-aware variance estimation.}
On top of the KM/NA linearization, we develop Greenwood-type plug-in formulas that separate ordinary KM variability from rank-mixing variability (e.g. Dell–Clutter misranking). These formulas are meant to be implementable in software and to show explicitly when RSS is delivering efficiency gains.

\para{3. Path to RSS log-rank / weighted log-rank.}
Because the rank-aware counting processes satisfy the same martingale CLTs, the same mechanics extend to two-sample log-rank and Fleming–Harrington–type tests; the present chapter builds the estimation theory needed for those later developments.

\section{Survival Estimation under Random Censoring and Ranked Set Sampling}

This section develops the asymptotic framework used in the rest of the paper. We start from the classical independent random‐censorship model and the product–integral representation of Kaplan–Meier (KM) and Nelson–Aalen (NA) estimators, then extend it to balanced ranked set sampling (RSS) with both perfect and imperfect (judgment) ranking. The key point is that, once each rank (true or judged) is treated as its own i.i.d.\ stratum with independent censoring, all of the usual SRS arguments (uniform laws, martingale representations, functional CLTs, Greenwood plug-ins) go through rank-wise and can then be averaged (cf.\ Supplements~\ref{supp:endpoints}--\ref{supp:rss-greenwood}).

\subsection{Baseline random–censorship model}

Let $X$ be the nonnegative lifetime and $C$ the nonnegative censoring time. We observe i.i.d.
\[
\{(Y_i,\delta_i): i=1,\dots,n\},\qquad
Y_i = \min(X_i,C_i), \quad \delta_i = I(X_i \le C_i),
\]
under the assumptions
(i) $(X_i,C_i)$ i.i.d., (ii) $X \perp C$, and (iii) $F(t)=P(X\le t)$, $G(t)=P(C\le t)$ are defined on $[0,\tau)$ with strictly positive observed‐time survival
\[
S_Y(t) := P(Y\ge t) = S(t)K(t) > 0 \quad\text{on } [0,\tau^\ast-\varepsilon],
\]
where $S=1-F$ and $K=1-G$, and $\tau^\ast$ is the right endpoint of $Y$ (see Supplement~\ref{supp:endpoints} for the distinction $\tau^\ast \le \tau$ and the interval of uniformity). The sub-distributions
\[
H_1(t) = P(Y\le t, \delta=1), \qquad
H_0(t) = P(Y\le t, \delta=0), \qquad
H = H_0 + H_1
\]
satisfy the standard identities $H_1(t) = \int_0^t K(u-)\,dF(u)$ and
\[
d\Lambda_1(u) = \frac{dH_1(u)}{S_Y(u-)} = \frac{dF(u)}{S(u-)}.
\]
These drive both the KM and NA estimators.

\subsection{Counting–process setup}

For each subject define the counting and at–risk processes
\[
N_i(t) = I(Y_i \le t, \delta_i = 1), \qquad Y_i(t) = I(Y_i \ge t),
\]
and their sums $N(t)=\sum_i N_i(t)$, $R(t)=\sum_i Y_i(t)$. With respect to the filtration generated by $\{Y_i(\cdot),N_i(\cdot)\}$, we have the Doob–Meyer decomposition
\[
N_i(t) = \int_0^t Y_i(u)\,d\Lambda_1(u) + M_i(t),
\]
with $M_i$ a square-integrable martingale. Summing gives
\[
N(t) = \int_0^t R(u)\,d\Lambda_1(u) + M(t),
\]
so the Nelson–Aalen estimator
\[
\widehat{\Lambda}_1(t) = \int_0^t \frac{dN(u)}{R(u)}
\]
admits the martingale transform representation
\[
\widehat{\Lambda}_1(t) - \Lambda_1(t) = \int_0^t \frac{1}{R(u)}\,dM(u),
\]
and on $[0,\tau^\ast-\varepsilon]$ we have $R(u)\ge c_\varepsilon n$ a.s.\ for all large $n$ (by a Glivenko–Cantelli argument; see Supplement~\ref{supp:positivity}), so $\widehat{\Lambda}_1 - \Lambda_1 = O_p(n^{-1/2})$.

The Kaplan–Meier estimator,
\[
\widehat{S}(t) = \prod_{u\le t} \Bigl(1 - \frac{dN(u)}{R(u)}\Bigr)
= \prod_{u\le t} (1 - d\widehat{\Lambda}_1(u)),
\]
is linked to NA through the product–integral map. A Taylor expansion of $\log(1-x)$ with $x=dN/R$ yields
\[
\log \widehat{S}(t) = - \widehat{\Lambda}_1(t) + r_n(t), \qquad
\sup_{t\le \tau^\ast-\varepsilon} |r_n(t)| = o_p(n^{-1/2}),
\]
so, at the CLT scale, KM and NA carry the same randomness (see Supplement~\ref{supp:prodint} for the full product–integral linearization).

\subsection{Empirical sub–distributions and strong law (SRS)}

Because the classes $\{I(Y\le t,\delta=d)\}$ are VC, the empirical sub-distributions
\[
H_{1n}(t) = \frac{1}{n}\sum_{i=1}^n I(Y_i\le t, \delta_i=1), \quad
H_{0n}(t) = \frac{1}{n}\sum_{i=1}^n I(Y_i\le t, \delta_i=0)
\]
converge uniformly a.s.\ to $(H_1,H_0)$ on $[0,\tau^\ast-\varepsilon]$. Thus $S_{Y,n} = 1 - H_n \to S_Y$ uniformly and $S_{Y,n}$ is bounded away from $0$ eventually. Substituting into
\[
\widehat{\Lambda}_1(t)
= \int_0^t \frac{dH_{1n}(u)}{S_{Y,n}(u-)},
\quad
\Lambda_1(t)
= \int_0^t \frac{dH_1(u)}{S_Y(u-)},
\]
and bounding the two “add–subtract’’ terms gives
\[
\sup_{t\le \tau^\ast-\varepsilon} |\widehat{\Lambda}_1(t) - \Lambda_1(t)| \xrightarrow{\text{a.s.}} 0,
\]
and, by continuity of the product–integral map,
\[
\sup_{t\le \tau^\ast-\varepsilon} |\widehat{S}(t) - S(t)| \xrightarrow{\text{a.s.}} 0.
\]
A detailed version of the add–subtract expansion is given in Supplement~\ref{supp:emp-srs}.

\subsection{Functional CLT and Greenwood variance (SRS)}

Applying Rebolledo’s martingale functional CLT (FCLT) to the stochastic integral
\(
\int_0^t R(u)^{-1} dM(u)
\)
and using $R(u)/n \to S_Y(u-)$ yields, in $\ell^\infty([0,\tau^\ast-\varepsilon])$,
\[
\sqrt{n}\,(\widehat{\Lambda}_1 - \Lambda_1) \Rightarrow \mathbb{G}_\Lambda,
\qquad
\mathrm{Cov}\bigl(\mathbb{G}_\Lambda(s),\mathbb{G}_\Lambda(t)\bigr)
= \int_0^{s\wedge t} \frac{dH_1(u)}{S_Y(u-)^2}.
\]
By the delta method for the product integral,
\[
\sqrt{n}\,(\widehat{S} - S) \Rightarrow \mathbb{G}_S := - S \,\mathbb{G}_\Lambda,
\quad
\mathrm{Cov}\bigl(\mathbb{G}_S(s),\mathbb{G}_S(t)\bigr)
= S(s)S(t)\int_0^{s\wedge t}\frac{dH_1(u)}{S_Y(u-)^2}.
\]
This gives, at any fixed $t$,
\[
\AVAR\bigl(\widehat{S}(t)\bigr)
= \frac{S(t)^2}{n} \int_0^t \frac{dH_1(u)}{S_Y(u-)^2} + o(n^{-1}),
\]
and the usual Greenwood plug-in
\[
\widehat{\AVAR}\bigl(\widehat{S}(t)\bigr)
= \widehat{S}(t)^2 \sum_{u\le t} \frac{dN(u)}{R(u)\bigl(R(u)-dN(u)\bigr)}
\]
is consistent (the “with-ties’’ version); see Supplement~\ref{supp:greenwood-srs} for an explicit derivation.

\subsection{Ranked set sampling with concomitants}

In balanced RSS we observe $n=mk$ units arranged as $k$ ranks over $m$ cycles. Under \emph{perfect} ranking, the lifetime among units measured at true rank $r$ is the $r$th order–statistic law, written $\Ftrue{r}$, with survival $\Strue{r}$ and observed-time survival $S_{Y,[r]}(t)=\Strue{r}(t)K(t)$. We compute a KM/NA curve \emph{within} each rank,
\[
\widehat{\Lambda}_{1,[r]}(t) = \int_0^t \frac{dN_r(u)}{R_r(u)}, \qquad
\widehat{S}_{[r]}(t) = \prod_{u\le t} \Bigl(1 - \frac{dN_r(u)}{R_r(u)}\Bigr),
\]
and estimate the population survival by the equal-weight average
\[
\widehat{S}_{\mathrm{RSS}}(t) := \frac{1}{k}\sum_{r=1}^k \widehat{S}_{[r]}(t).
\]
McIntyre’s RSS identity $\frac{1}{k}\sum_{r=1}^k \Strue{r}(t)=S(t)$ then gives
\(
\widehat{S}_{\mathrm{RSS}} \to S
\)
uniformly on $[0,\tau^\ast-\varepsilon]$ as $m\to\infty$. The rank-wise i.i.d.\ structure and independence across ranks (because each cycle uses $k$ independent candidate sets) are spelled out in Supplement~\ref{supp:rss-iid}.

In practice, ranking is done on a proxy $\tilde X = X + Z$ (Dell–Clutter model), so we work with \emph{judged} ranks $J=r$. Section~2.2 above showed that the judged-$r$ lifetime cdf is a mixture
\[
F_r^J(t) = \sum_{j=1}^k w_{rj} F_{[j]}(t),\qquad
w_{rj}=P(T=j\mid J=r), \ \sum_j w_{rj}=1,
\]
and that, under balanced selection $P(J=r)=1/k$ and uniform true-rank frequencies $P(T=j)=1/k$, the average over $r$ recovers the population:
\[
\frac{1}{k}\sum_{r=1}^k F_r^J(t) = F(t), \qquad
\frac{1}{k}\sum_{r=1}^k S_r^J(t) = S(t).
\]
The conditioning proof of this identity is given in Supplement~\ref{supp:mixing}. Hence the \emph{judgment-rank} KM,
\[
\widehat{S}_{\mathrm{RSS},J}(t) := \frac{1}{k}\sum_{r=1}^k \widehat{S}^{J}_{r}(t),
\]
is again a consistent estimator of $S(t)$.

\subsection{CLT and efficiency ordering under RSS}

Because each rank contributes $m$ i.i.d.\ observations, the SRS martingale FCLT applies \emph{within} rank:
\[
\sqrt{m}\bigl(\widehat{S}_{[r]} - S_{[r]}\bigr) \Rightarrow \mathbb{G}_{S,[r]},
\qquad
\mathrm{Cov}(\mathbb{G}_{S,[r]}(s),\mathbb{G}_{S,[r]}(t))
= S_{[r]}(s)S_{[r]}(t)\int_0^{s\wedge t} \frac{dH_{1,[r]}(u)}{S_{Y,[r]}(u-)^2}.
\]
With $n=mk$ and independence across ranks,
\[
\sqrt{n}\bigl(\widehat{S}_{\mathrm{RSS}} - S\bigr)
= \frac{1}{\sqrt{k}} \sum_{r=1}^k \sqrt{m}\bigl(\widehat{S}_{[r]} - S_{[r]}\bigr)
\Rightarrow \mathbb{G}^{(\mathrm{perf})}_S
\]
with covariance kernel
\[
\mathrm{Cov}\bigl(\mathbb{G}^{(\mathrm{perf})}_S(s),\mathbb{G}^{(\mathrm{perf})}_S(t)\bigr)
= \frac{1}{k} \sum_{r=1}^k
S_{[r]}(s)S_{[r]}(t)\int_0^{s\wedge t}\frac{dH_{1,[r]}(u)}{S_{Y,[r]}(u-)^2}.
\]
The judgment-rank limit is identical except that $[r]$ is replaced by $J=r$ and the rank-specific laws are mixtures (see Supplement~\ref{supp:emp-rss} for the RSS add–subtract expansion mirroring the SRS one). Comparing the SRS kernel
\[
S(s)S(t)\int_0^{s\wedge t}\frac{dH_1(u)}{S_Y(u-)^2}
\]
with the two RSS kernels yields the pointwise ordering
\[
V_{\mathrm{perf}}(t) \ \le\ V_{\mathrm{judg}}(t;\rho) \ \le\ V_{\mathrm{SRS}}(t),
\]
with equalities at $\rho=1$ (perfect ranking) and $\rho=0$ (uninformative ranking). Thus, rank information never hurts first-order efficiency; it helps whenever the proxy is even moderately informative.

\subsection{Variance estimation for rank-aware KM}

Because the RSS KM is an equal-weight average of $k$ independent KM curves, the natural plug-ins are simply “sum of Greenwoods, divided by $k^2$’’:
\[
\widehat{\Var}\{\widehat{\Lambda}_{\mathrm{RSS}}(t)\}
= \frac{1}{k^2}\sum_{r=1}^k \sum_{u\le t} \frac{dN_r(u)}{R_r(u)^2},
\quad
\widehat{\Var}\{\widehat{S}_{\mathrm{RSS}}(t)\}
= \frac{1}{k^2}\sum_{r=1}^k \widehat{S}_{[r]}(t)^2
\sum_{u\le t} \frac{dN_r(u)}{R_r(u)\{R_r(u)-dN_r(u)\}}.
\]
The same formulas work for judgment ranks with $\widehat{S}^{J}_r$ and $H^{J}_{1,r}$, and they converge to the kernels in the RSS CLTs above (see Supplement~\ref{supp:rss-greenwood}).

\subsection{Simulation study}
\label{sec:sim}

We assess the finite-sample behavior of these estimators under two superpopulations, chosen to match the theory as closely as possible while still producing appreciable rankwise heterogeneity.

\para{(i) Log–AFT superpopulation:}
We generate lifetimes from
\[
X = \exp(\mu - \beta Z + \varepsilon), \qquad
Z\sim N(0,1),\ \varepsilon \sim N(0,\sigma_\varepsilon^2),
\]
with $(Z,\varepsilon)$ independent and baseline values $\mu=0$, $\beta=1.5$, $\sigma_\varepsilon=0.4$. A large i.i.d.\ sample from this model is first generated to (a) fix four evaluation times at survival levels
\[
S(t_{0.25}) \approx 0.75,\quad S(t_{0.50}) \approx 0.50,\quad
S(t_{0.75}) \approx 0.25,\quad S(t_{0.90}) \approx 0.10,
\]
and (b) calibrate ranking noise. To mimic practical RSS, we rank on a noisy concomitant
\[
\tilde{X} = Z + U,\qquad U \sim N(0,\sigma_{\tilde{X}}^2),
\]
and choose $\sigma_{\tilde{X}}$ numerically so that $|\Corr(\tilde{X},X)|$ is close to a prescribed $\rho_{\text{target}}\in\{0.1,0.3,0.5,0.7,0.9\}$. This is the exact analogue of the Dell–Clutter $\rho=\Corr(\tilde X,X)$ used in the theory.

Balanced RSS samples are then generated exactly as in the model: for each cycle we create $k$ independent candidate sets of size $k$, rank within the $r$-th set by $W$ and \emph{measure only} the unit judged to be rank $r$, $r=1,\dots,k$. Repeating over $m$ cycles yields $n=mk$ observations and $m$ observations per judged rank, so the rank-aware KM is just
\(
\widehat{S}_{\mathrm{RSS}}(t) = \frac{1}{k}\sum_{r=1}^k \widehat{S}^{J}_r(t).
\)

Censoring is imposed independently by $C\sim \mathrm{Exp}(-\log(1-p_{\text{cens}})/\mathbb{E}X)$ with $p_{\text{cens}}\in\{0,0.1,0.3,0.5\}$, and the \emph{same} censoring law is used for the SRS benchmark of size $n=mk$. For each replicate we compute:
(i) the SRS KM,
(ii) the rank-aware RSS KM, and
(iii) the corresponding Greenwood plug-ins from the previous subsection.

The grid is
\[
k \in \{2,4,6,8,10\},\quad
m \in \{20,50\},\quad
\rho_{\text{target}} \in \{0.1,0.3,0.5,0.7,0.9\},\quad
p_{\text{cens}} \in \{0,0.1,0.3,0.5\}.
\]
For every cell we run $B_{\text{MC}}=10{,}000$ Monte Carlo replicates, which keeps the relative Monte Carlo error of variance estimates around $1.4\%$ and makes variance \emph{ratios}
\[
\RE_{\text{MC}}(t) =
\frac{\widehat V_{\text{SRS,MC}}(t)}{\widehat V_{\text{RSS,MC}}(t)}, \qquad
\RE_{\text{GW}}(t) =
\frac{\overline{\widehat{\Var}(\widehat S_{\text{SRS}}(t))}}{\overline{\widehat{\Var}(\widehat S_{\text{RSS}}(t))}},
\]
stable. A secondary run with $B_{\text{true}}=4{,}000$ replicates is used only to smooth the “theoretical’’ ratio
\(
\RE_{\text{true}}(t) = V_{\text{SRS,true}}(t)/V_{\text{RSS,true}}(t)
\)
over the grid. Because the SRS and RSS samples in a replicate always share $(n,p_{\text{cens}})$, the efficiency comparison is fair.

Here we present the simulation results graphically to show how the relative efficiency changes with evaluation time, $k$, $m$, $\rho$ and censoring level.



\begin{figure}[p]
	\centering
	\begingroup
	\def\AFTpath{}%
	\captionsetup{aboveskip=4pt, belowskip=0pt}
	\captionsetup[subfigure]{justification=centering,aboveskip=2pt,belowskip=2pt}
	\setlength{\textfloatsep}{6pt}\setlength{\floatsep}{6pt}\setlength{\intextsep}{6pt}
	
	\begin{subfigure}{\textwidth}\centering
		\includegraphics[width=\textwidth,height=0.30\textheight,keepaspectratio]{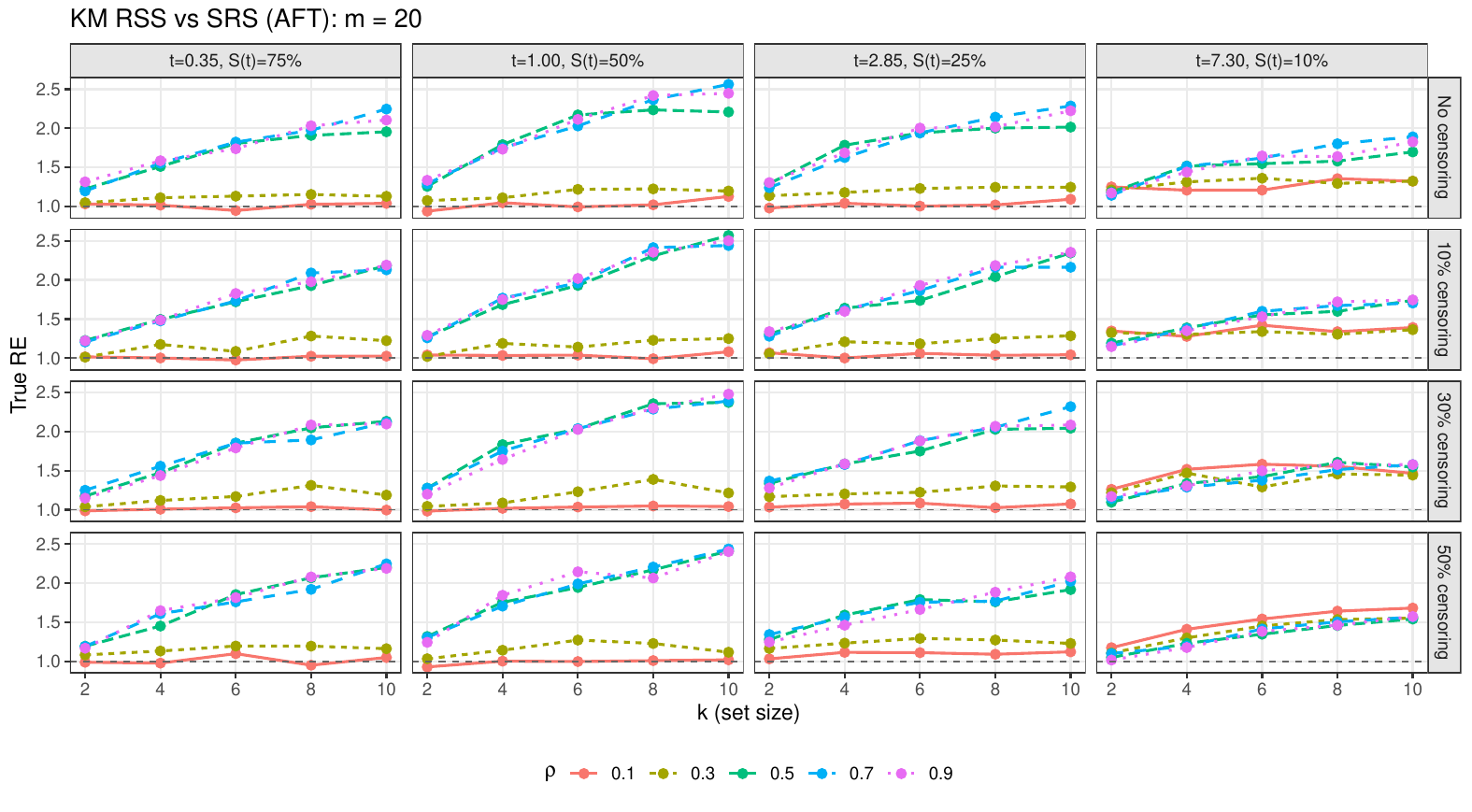}
		\caption{True RE, $m=20$}
	\end{subfigure}
	
	\vspace{0.25em}
	
	\begin{subfigure}{\textwidth}\centering
		\includegraphics[width=\textwidth,height=0.30\textheight,keepaspectratio]{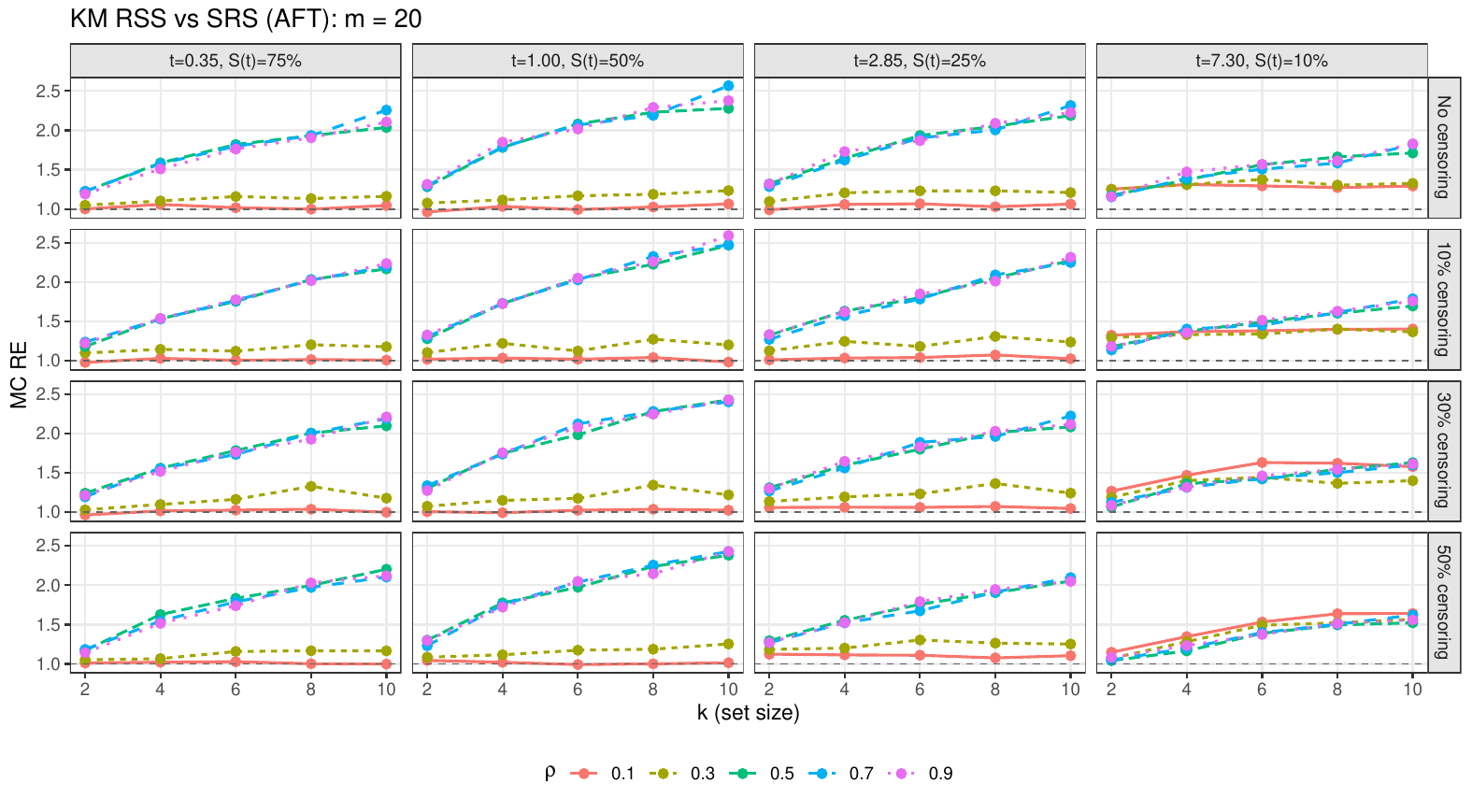}
		\caption{MC RE, $m=20$}
	\end{subfigure}
	
	\vspace{0.25em}
	
	\begin{subfigure}{\textwidth}\centering
		\includegraphics[width=\textwidth,height=0.30\textheight,keepaspectratio]{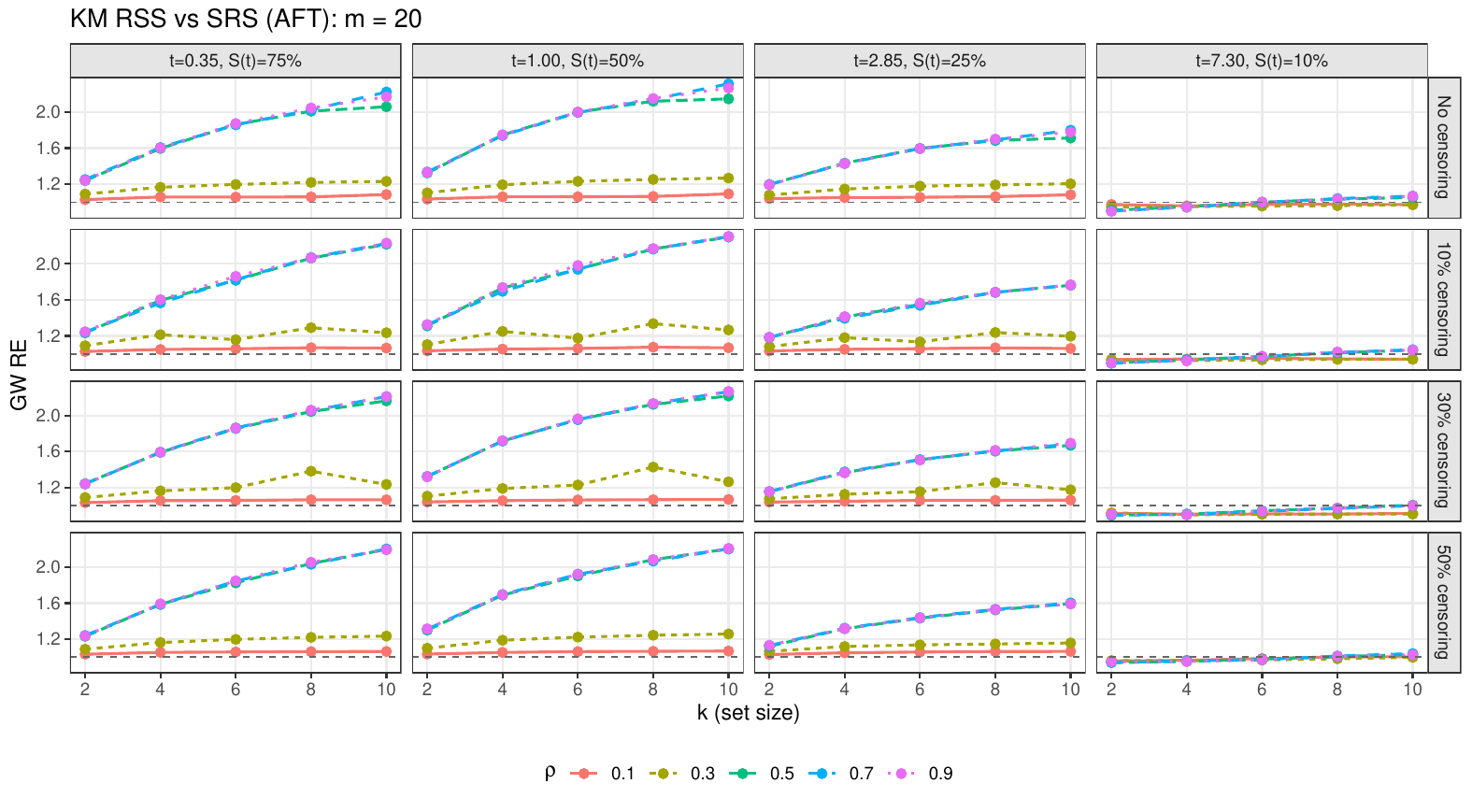}
		\caption{GW RE, $m=20$}
	\end{subfigure}
	
	\caption{AFT superpopulation: relative efficiency (RSS vs.\ SRS) across set size $k$,
		ranking quality $\rho$ (legend), censoring fractions (rows), and evaluation times (columns).
		Dashed line marks RE$=1$.}
	\label{fig:aft-m20-tall}
	\endgroup
\end{figure}
\clearpage

\begin{figure}[p]
	\centering
	\begingroup
	\def\AFTpath{}%
	\captionsetup{aboveskip=4pt, belowskip=0pt}
	\captionsetup[subfigure]{justification=centering,aboveskip=2pt,belowskip=2pt}
	\setlength{\textfloatsep}{6pt}\setlength{\floatsep}{6pt}\setlength{\intextsep}{6pt}
	
	\begin{subfigure}{\textwidth}\centering
		\includegraphics[width=\textwidth,height=0.30\textheight,keepaspectratio]{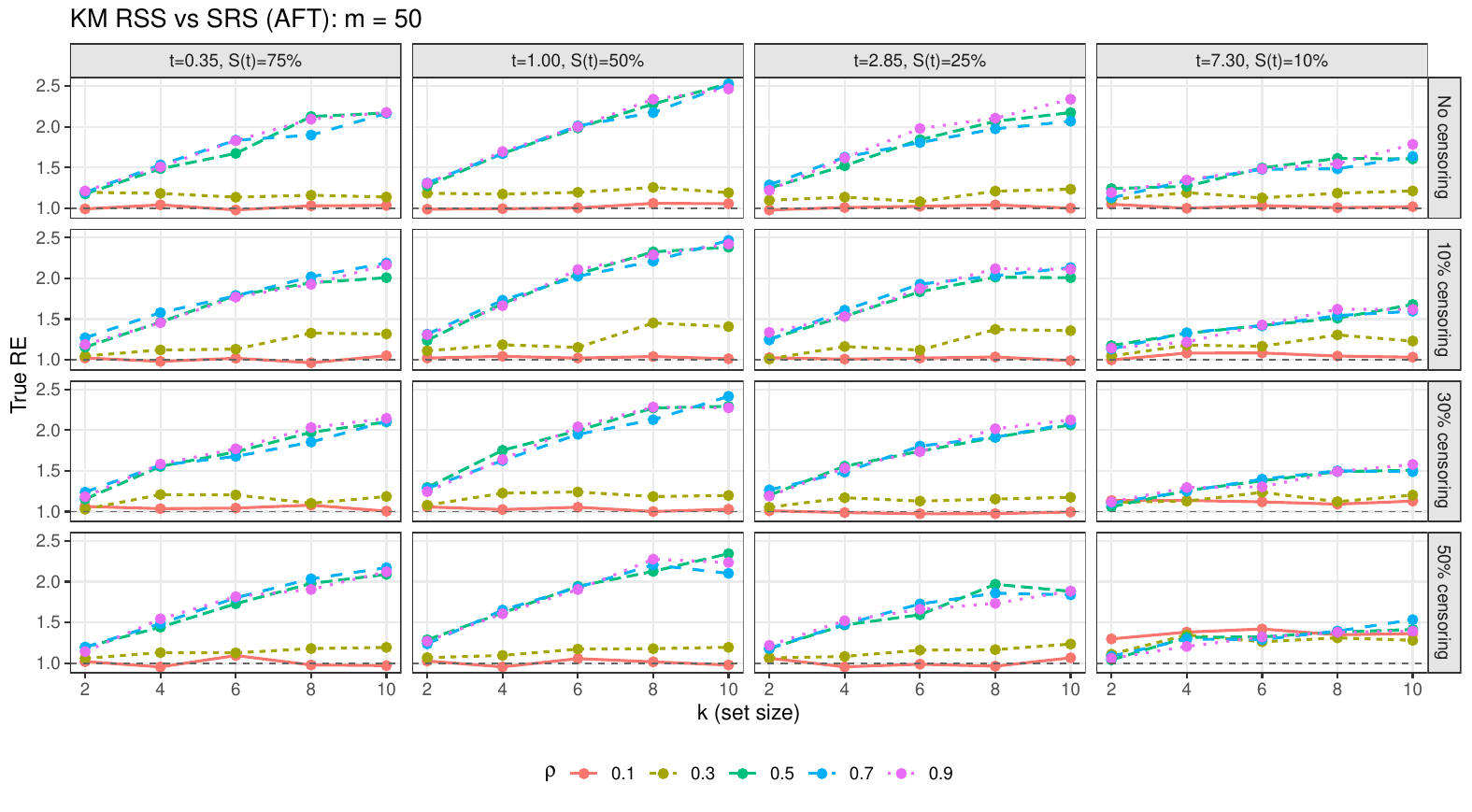}
		\caption{True RE, $m=50$}
	\end{subfigure}
	
	\vspace{0.25em}
	
	\begin{subfigure}{\textwidth}\centering
		\includegraphics[width=\textwidth,height=0.30\textheight,keepaspectratio]{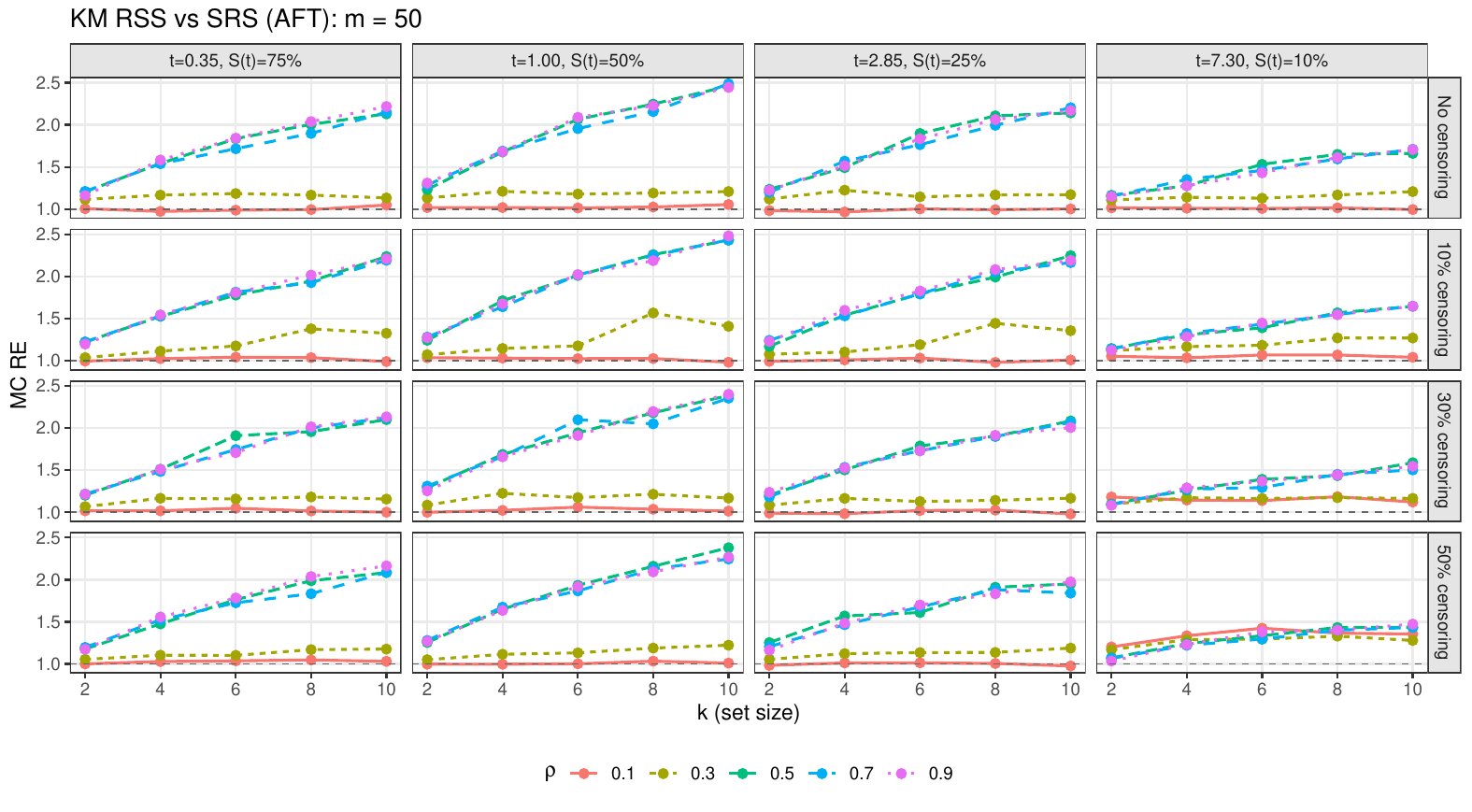}
		\caption{MC RE, $m=50$}
	\end{subfigure}
	
	\vspace{0.25em}
	
	\begin{subfigure}{\textwidth}\centering
		\includegraphics[width=\textwidth,height=0.30\textheight,keepaspectratio]{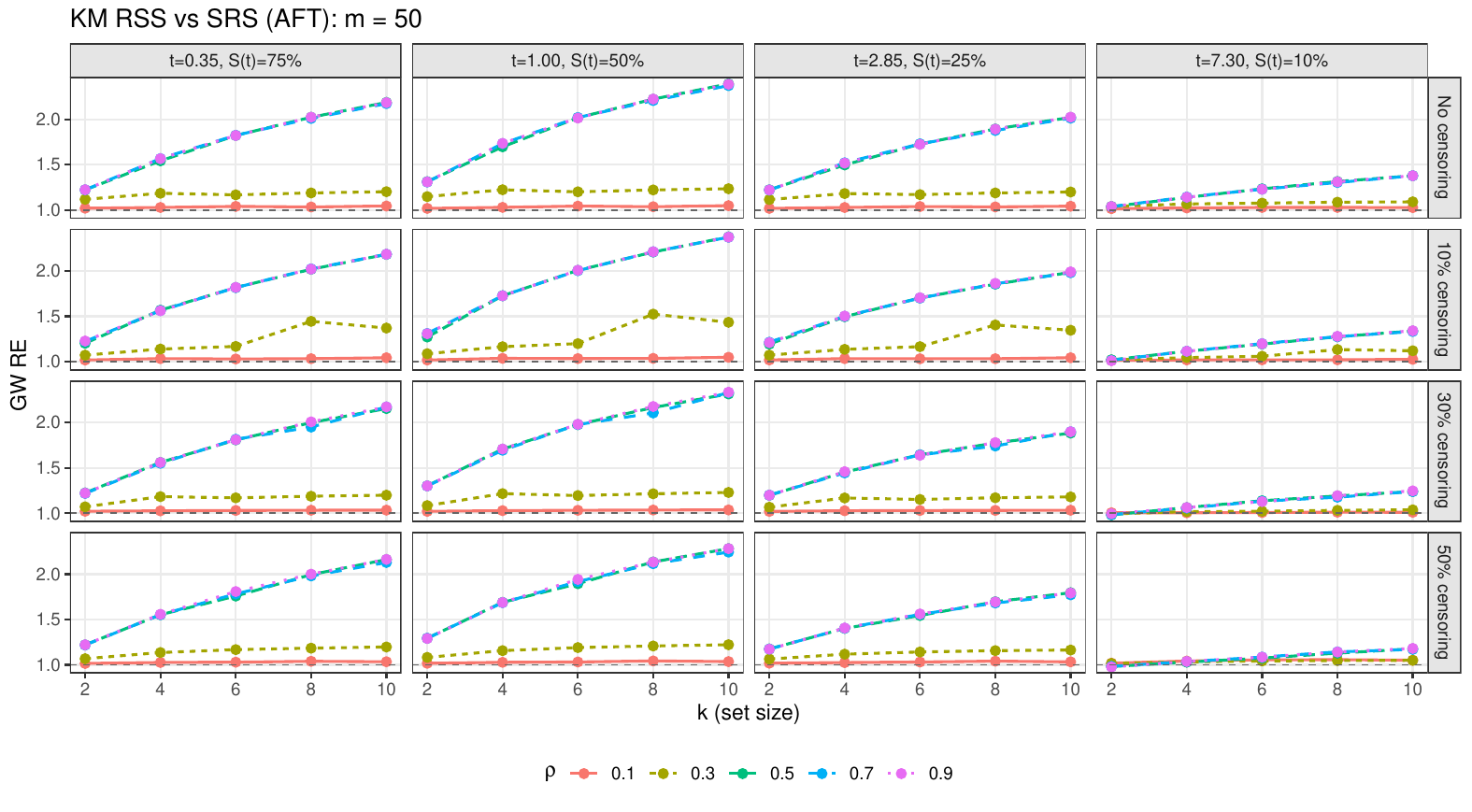}
		\caption{GW RE, $m=50$}
	\end{subfigure}
	
	\caption{AFT superpopulation: relative efficiency (RSS vs.\ SRS) with $m=50$ cycles. Layout as in Fig.~\ref{fig:aft-m20-tall}.}
	\label{fig:aft-m50-tall}
	\endgroup
\end{figure}
\clearpage

\para{Main findings under the AFT superpopulation}

Across all panels the same pattern repeats: relative efficiency (RSS vs.\ SRS) increases monotonically with ranking quality $\rho$ and with set size $k$. Even with only moderately informative ranks ($\rho\approx 0.5$), choosing $k\in\{6,8,10\}$ yields clear gains—typically $\mathrm{RE}\approx 1.2$–$1.6$ at the first two evaluation times—while strong ranking ($\rho\approx 0.9$) pushes the gains toward a factor of two as $k$ grows. The evaluation time matters: improvements are largest when we evaluate earlier on the survival curve ($S(t)=75\%$ and $50\%$), taper at $S(t)=25\%$, and are smallest at $S(t)=10\%$, where risk sets are tiny and late–time variability dominates, especially under censoring. Increasing the censoring fraction shifts every curve downward but does not erase the advantage of RSS; for $\rho\ge 0.5$ and $k\ge 6$ the gains remain visible at the first three time points even with $30$–$50\%$ censoring. Comparing $m=20$ with $m=50$ shows almost no qualitative change in the RE profiles, which is consistent with the theory that, at fixed total size $n=mk$, first–order efficiency is driven by how $n$ is allocated across ranks (via $k$) and by ranking quality (via $\rho$), not by the number of cycles $m$. The three variance ratios agree well: the Monte Carlo ratio $\mathrm{RE}_{\mathrm{MC}}$ closely tracks the larger-run benchmark $\mathrm{RE}_{\mathrm{true}}$ across times and censoring levels, and the Greenwood plug-in ratio $\mathrm{RE}_{\mathrm{GW}}$ aligns with both at the first three time points. At $S(t)=10\%$, however, $\mathrm{RE}_{\mathrm{GW}}$ deteriorates, reflecting the well-known instability of the classical Greenwood variance when risk sets are small and ties are present; this motivates a tail-robust plug-in in future work. Overall, the figures confirm the theoretical ordering—better ranking and larger $k$ yield smaller variance for the rank-aware KM—and show that practically meaningful gains arise already for moderate $\rho$ and sample sizes.

\para{(ii) Weibull superpopulation:}
To show that the same ordering is visible in a fully parametric survival family, we repeat an smaller simulation with exponential lifetimes
\[
X \sim \text{Weibull}(\nu=1,\theta_1=1), \quad S(t)=e^{-t},
\]
and independent Weibull censoring with the same shape $\nu$ and scale
\(
\theta_2 = \bigl(\frac{1-p_{\text{cens}}}{p_{\text{cens}}}\bigr)^{1/\nu}
\)
to hit $p_{\text{cens}}\in\{0,0.1,0.3,0.5\}$. Evaluation times are again chosen at survival levels $0.75,0.50,0.25,0.10$, which here are just $t=-\log(0.75),\dots,-\log(0.10)$.

Imperfect ranking is introduced by the Dell–Clutter rule
\[
\widetilde X = X + Z,\qquad Z\sim N(0,\sigma_Z^2),
\]
with $\sigma_Z^2$ chosen from the closed-form Dell–Clutter relation
\(
\sigma_Z^2 = \Var(X)\bigl(\rho^{-2}-1\bigr)
\)
to get $\rho\in\{0.1,0.3,0.5,0.7,0.9\}$. Sampling is the same balanced RSS as above. We use the same grid
\(
k\in\{4,6,8,10\}, m\in\{20,50\}, \rho\in\{0.1,\dots,0.9\}, p_{\text{cens}}\in\{0,0.1\}
\)
and $B_{\text{MC}}=1{,}000$ replicates. In this Weibull case we can also compute the asymptotic KM variances analytically from the kernels in the previous subsections, so we report the asymptotic relative efficiencies; in our runs these track closely, confirming that the RSS variance reduction is not an artifact of the AFT generator.

\begin{figure}[h!]
	\centering
	\captionsetup{skip=4pt} 
	\includegraphics[width=\textwidth,keepaspectratio]{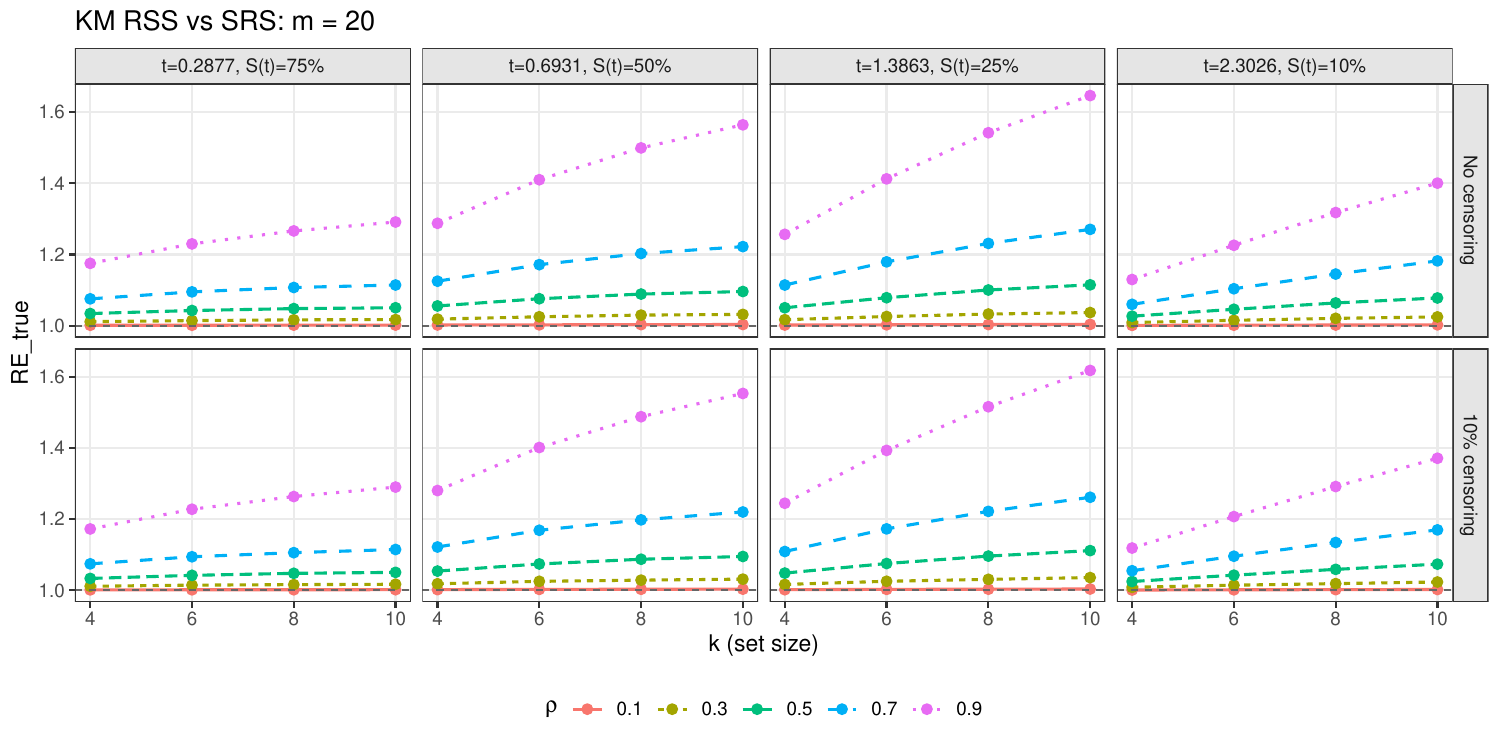}
	\caption{Weibull superpopulation (exponential baseline). \textbf{True RE} (RSS vs.\ SRS) across set size $k$, ranking quality $\rho$ (legend), censoring fractions (rows), and evaluation times (columns); dashed line marks RE = 1. $m = 20$ cycles.}
	\label{fig:weibull-re-m20}
\end{figure}

\begin{figure}[h!]
	\centering
	\captionsetup{skip=4pt}
	\includegraphics[width=\textwidth,keepaspectratio]{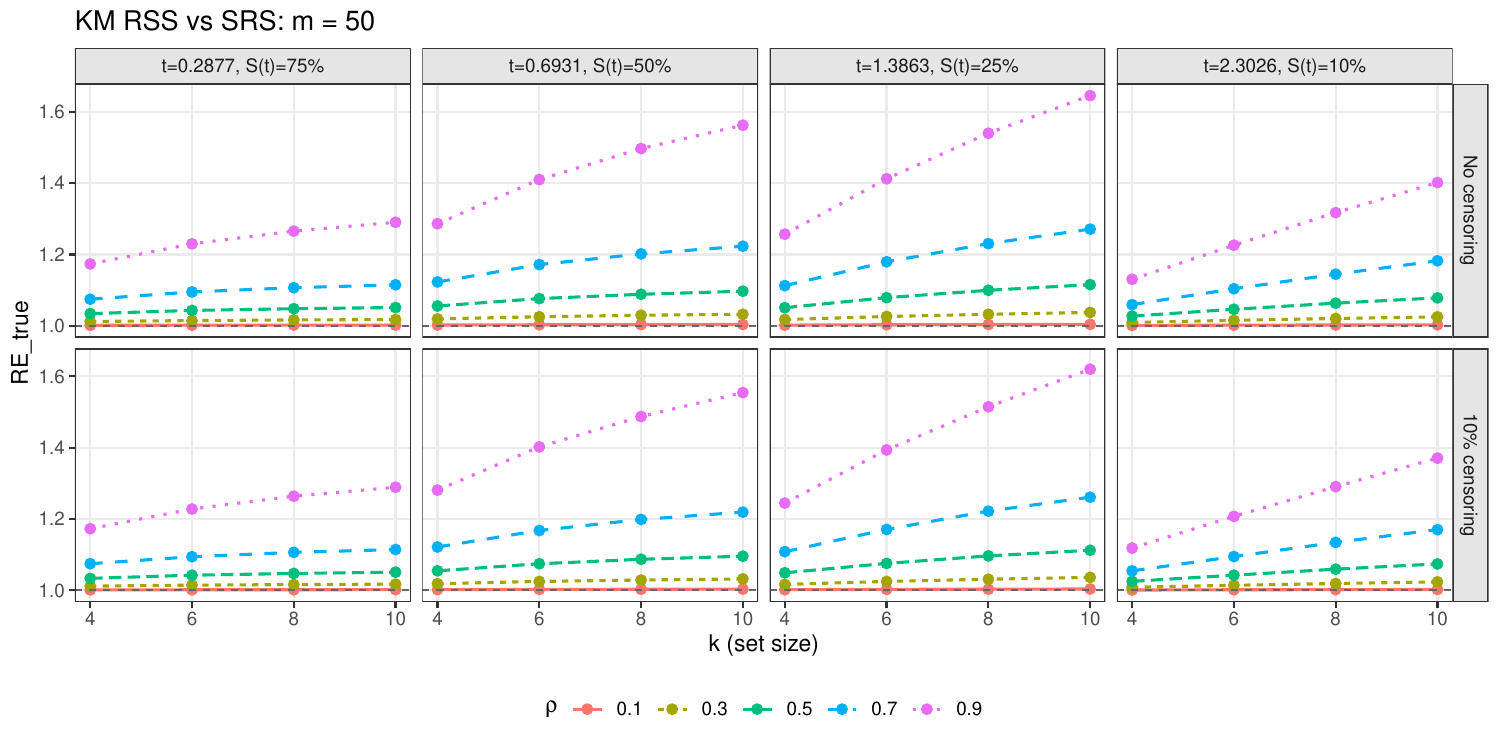}
	\caption{Weibull superpopulation (exponential baseline). \textbf{True RE} (RSS vs.\ SRS) across set size $k$, ranking quality $\rho$ (legend), censoring fractions (rows), and evaluation times (columns); dashed line marks RE = 1. $m = 50$ cycles.}
	\label{fig:weibull-re-m50}
\end{figure}

\subsection{Contributions and novelty}

Relative to the existing literature on KM under censoring \cite{Stute1993,Stute1995} and on RSS with survival outcomes \cite{StrzalkowskaKominiak2014,mahdizadeh2017resampling}, this section contributes four points.

\begin{enumerate}
	\item We give \emph{process-level} (not just functional) CLTs for the KM and NA estimators on $[0,\tau-\varepsilon]$ under independent random censoring, stated directly in terms of $H_1$ and $S_Y$, and we do so via the counting–process/product–integral route of \cite{AndersenBorganGillKeiding1993} and the Hadamard differentiability of the product integral \cite{GillJohansen1990}. This makes the linearization automatically tie-robust and produces Greenwood-type plug-ins compatible with discrete event times.
	
	\item We extend these results to \emph{balanced RSS with perfect ranking}, deriving strong laws and functional CLTs both within rank and for the equal-weight average, and we write the limiting covariance as the simple average of the $k$ within-rank kernels. This delivers the process-level theory needed for simultaneous bands and time-indexed tests for RSS KMs, which was missing from earlier, mainly simulation-based, work.
	
	\item We treat \emph{imperfect (judgment) ranking} explicitly via the Dell–Clutter concomitant model, push the induced true/judged mixing through the NA/KM machinery, and obtain LLNs, FCLTs, and variance kernels that depend monotonically on a ranking-quality parameter $\rho\in[0,1]$. Within this framework we prove the efficiency ordering
	\[
	V_{\mathrm{perf}}(t) \le V_{\mathrm{judg}}(t;\rho) \le V_{\mathrm{SRS}}(t)
	\]
	with equality only at $\rho=1$ and $\rho=0$, using conditional-variance (tower) arguments applied to the KM influence function.
	
	\item We spell out the exact target when the true-rank distribution of measured units is not uniform: the rank average estimates
	\(
	S^\star(t) = \sum_j P(T=j) S_{[j]}(t),
	\)
	which coincides with $S(t)$ only when $P(T=j)=1/k$. We also indicate precisely where rank-invariant censoring and the independence across ranks/cycles are used. A practical corollary is that, at fixed $n=mk$, the first-order relative efficiency depends on $k$ and on ranking quality $\rho$, but not on $m$.
\end{enumerate}
	
\section{Future Work and Planned Extensions}
\label{sec:future-work}

The results in this paper establish the rank-aware KM/NA theory for balanced RSS (perfect and judgment ranking), provide plug-in Greenwood estimators, and show—via simulations—that the variance ordering predicted by the theory can be observed in finite samples. Several natural extensions follow from the same tools but are left for future work.

\subsection{Developing a tail-robust variance for the rank-aware KM}
The AFT figures indicate that rank-average Greenwood inflates at late times where per-rank risk sets become small, so we will stabilize inference while retaining the estimator $\widehat S_{\mathrm{RSS}}(t)=k^{-1}\sum_r \widehat S_r(t)$. First, we will use monotone variance-stabilizing transforms—most notably the complementary log--log $\theta(t)=\log\{-\log \widehat S_{\mathrm{RSS}}(t)\}$ (and, if needed, logit or arcsine–square-root)—apply delta-method standard errors on the transformed scale, and back-transform for reporting to improve tail calibration. Second, we will compute a pooled-risk-set Greenwood (forming $R_{\text{tot}}$ by ignoring ranks) as a conservative lower bound and shrink the rank-average Greenwood toward it with a modest weight, yielding a stable sandwich between pooled (lower) and rank-average (upper) variances without altering the point estimator. Third, we will implement a rank-wise multiplier (perturbation) bootstrap that reweights the counting- and at-risk processes within each rank to approximate the finite-sample distribution of $\widehat S_{\mathrm{RSS}}(t)$ even when some $R_r(u)$ are small. These steps directly target the observed tail behavior while remaining consistent with first-order theory on $[0,\tau^\ast-\varepsilon]$.

\para{Pooled-risk-set shrinkage (intuition)}

Late in follow-up, each judged rank carries only a few individuals at risk, so the rank-wise Greenwood increments become very “spiky’’ and, when averaged across ranks, tend to overstate uncertainty simply because we are averaging over several small denominators. If, instead, we momentarily ignore the rank labels and treat all units as one pooled risk set, the same Greenwood construction is much more stable and, by design, provides a conservative lower bound on the variance near the tail. Our plan is to blend these two signals: keep the rank-aware estimator of $S(t)$ exactly as is, but shrink its rank-average Greenwood variance toward the pooled-risk variance by a modest weight that increases only when information visibly thins out (i.e., when effective risk sets are small). This should produce a variance that stays between a stable lower bound from pooling and the usual rank-average estimate that works well away from the tail, curbs tail inflation without changing the point estimator, and remains aligned with the first-order theory on $[0,\tau^\ast-\varepsilon]$.

\para{Rank-wise multiplier bootstrap}

To approximate the finite-sample law of $\widehat S_{\mathrm{RSS}}(t)$ on the \emph{observed} dataset without re-drawing subjects, we use a perturbation bootstrap that respects the RSS strata. Within each judged rank $r$ (and cycle index $j$), draw i.i.d.\ \emph{nonnegative} weights $W_{rj}$ with mean $1$ and variance $1$ (e.g., $W_{rj}\sim\mathrm{Exp}(1)$ or, more generally, $\mathrm{Gamma}$ weights). Form weighted counting and risk processes
\[
N_r^{\ast}(t)=\sum_{j=1}^m W_{rj}\,I(Y_{rj}\le t,\delta_{rj}=1),
\qquad
R_r^{\ast}(t)=\sum_{j=1}^m W_{rj}\,I(Y_{rj}\ge t),
\]
compute the weighted NA/KM within rank and average $\widehat S_{\mathrm{RSS}}^{\ast}(t)=k^{-1}\sum_{r=1}^k \widehat S_r^{\ast}(t)$. Repeating this $B^\ast$ times yields $\widehat{\Var}_\ast\{\widehat S_{\mathrm{RSS}}(t)\}$ from the sample variance across replicates. Using positive, mean-one weights preserves the at-risk structure and aligns with the counting-process martingale representation, providing a stable tail behavior; a within-rank resampling bootstrap is also possible but is computationally heavier and can be less stable when risk sets are small.

\subsection{Rank-aware two-sample tests}
A second direction is to develop RSS versions of the standard two-sample survival tests, beginning with the log–rank test and continuing with weighted log–rank families (e.g. Fleming–Harrington–type weights). The idea is to replace the SRS counting and at–risk processes in the usual “observed minus expected” statistic by their rank-wise RSS analogues and to average the corresponding variance contributions across ranks, just as we did for the KM variance. This will require (i) writing the RSS numerator as a sum of rank-level martingale terms, (ii) identifying the predictable variation under the null when the two groups share the same rank-specific hazards, and (iii) checking that imperfect (judgment) ranking only inflates the variance toward the SRS limit, in the same way we proved for KM. Doing so would make RSS usable in the common “group A vs group B” survival comparisons.

\subsection{Functional estimands: RML and WML}
A third direction is to extend the rank-aware CLTs to time-integrated functionals such as the restricted mean life (RML) and window mean life (WML). Since our main results already give a pooled, rank-averaged KM with a closed-form covariance kernel on $[0,\tau^\ast-\varepsilon]$, the next step is to integrate that process over time (using the usual step or trapezoidal rule on the observed event-time grid) and apply a delta-method argument to obtain asymptotic normality of the RML/WML estimators. This would allow point estimates and Wald-type intervals/tests for integrated survival under RSS, which is important when survival curves differ mostly early or mostly late and KM-based pointwise CIs are hard to interpret.

\subsection{Design guidance and determination of sample size}
A fourth direction is to turn the simulation patterns in Section~\ref{sec:sim} into explicit guidelines. The empirical results suggest that, at fixed $n=mk$,
\begin{itemize}
	\item increasing the set size $k$ gives a clearer efficiency gain than increasing the number of cycles $m$;
	\item improving the concomitant (i.e. getting a larger correlation/ranking quality) is often more beneficial than collecting more cycles with a weak concomitant;
	\item under heavy censoring, inference should be restricted to $[0,\tau^\ast-\varepsilon]$ and the chosen $\varepsilon$ should be reported.
\end{itemize}
The future work will formalize these statements as simple “design recipes” (e.g. recommended $k$ in the 4–8 range; minimum $m$ for stable Greenwood SEs; how to diagnose when a concomitant is too weak to justify RSS).

\subsection{Applications to large public datasets}
To demonstrate feasibility in practice, we plan to pick two large, publicly available datasets and treat each of them as a superpopulation from which we can repeatedly draw artificial SRS and RSS samples to display the efficiency gain.

\para{Reliability-style dataset}

Candidates include the Backblaze drive-failure tables (daily drive-level records). A natural survival outcome is “time from first appearance in the log to drive failure”, with censoring at the last day the drive is seen. Plausible concomitants for ranking inside each $k$-set are drive age, early SMART anomalies, or a simple composite risk score. Natural groupings (for future RSS log–rank/WML tests) are model family, capacity tier, or HDD vs SSD.

\para{Medical/patient-outcome dataset}

Candidates include SEER-style cancer registries (time-to-death, censored at last follow-up) with rich baseline variables: stage, summary stage, tumor size bands, nodal status, histology/grade, and treatment-at-diagnosis indicators (surgery, radiation, chemo). These variables can act as concomitants for ranking (more advanced disease $\Rightarrow$ earlier expected event) and as grouping variables for comparisons. A second medical candidate is TCGA ovarian (OV), where clinical tables typically contain days-to-death or days-to-last-follow-up, residual disease / cytoreduction status, stage, and treatment response; residual disease or debulking status can serve as an informative concomitant for ranking patients inside each $k$-set, while stage or therapy response can define the groups to be compared. Leaving the exact pair of datasets open will let the committee recommend the most compelling combination (e.g. one reliability, one clinical).

\subsection{Software}
Finally, we plan to release an R package that (i) takes data in the $(Y,\delta,J)$ layout used in this paper, (ii) computes SRS and rank-aware RSS KM/NA side by side, (iii) produces the Greenwood plug-ins for both, and (iv) includes simulators that can draw balanced RSS samples from a large “population table” (such as the datasets above). This will make it easy for applied researchers to try RSS-based survival estimation without re-implementing the rank-aware pieces.

\printbibliography

\newpage
\appendix
\section{Technical Supplements}

\subsection{Endpoints and interval of uniformity}
\label{supp:endpoints}
Let
\[
\tau := \sup\{t \ge 0 : S(t) > 0\}, \qquad
\tau^\ast := \sup\{t \ge 0 : S_Y(t) > 0\},
\]
where $S(t)=P(X>t)$ and $S_Y(t)=P(Y\ge t)=S(t)K(t)$. Because $Y=\min(X,C)$, we always have
$\tau^\ast \le \tau$. All uniform laws and functional CLTs in the paper are stated on
$[0,\tau^\ast-\varepsilon]$ for an arbitrary $\varepsilon>0$ so that $S_Y$ is bounded away from $0$ and
the at–risk process can be inverted.

\subsection{Uniform positivity of the at–risk process}
\label{supp:positivity}
Fix $\varepsilon>0$. The class $\{I(Y \le t): t \in [0,\tau^\ast-\varepsilon]\}$ is a VC class, hence
\[
\sup_{t \le \tau^\ast-\varepsilon} |H_n(t) - H(t)| \xrightarrow{\text{a.s.}} 0.
\]
Since $S_{Y,n}(t)=1-H_n(t)$ and $S_Y(t)=1-H(t)$, we obtain
\[
\sup_{t \le \tau^\ast-\varepsilon} |S_{Y,n}(t) - S_Y(t)| \xrightarrow{\text{a.s.}} 0.
\]
Choose $c_\varepsilon>0$ with $\inf_{t \le \tau^\ast-\varepsilon} S_Y(t) \ge 2c_\varepsilon$. For all large $n$,
\[
\inf_{t \le \tau^\ast-\varepsilon} S_{Y,n}(t) \ge c_\varepsilon.
\]
Because $R(t)=n S_{Y,n}(t)$, this gives
\[
R(t) \ge c_\varepsilon n \qquad \text{for all } t \le \tau^\ast-\varepsilon,
\]
which is the bound used in the counting–process representation in the main text.

\subsection{Product–integral linearization and KM remainder}
\label{supp:prodint}
Let $\Lambda_n \to \Lambda$ uniformly on $[0,\tau^\ast-\varepsilon]$, with $\Lambda$ of bounded
variation. Define
\[
S_n(t) = \prod_{u \le t} (1 - d\Lambda_n(u)), \qquad
S(t) = \prod_{u \le t} (1 - d\Lambda(u)).
\]
By the Hadamard differentiability of the product integral
(see Gill and Johansen, 1990, and Fleming and Harrington, 1991),
\[
\log S_n(t) - \log S(t)
= -\{\Lambda_n(t) - \Lambda(t)\} + o\bigl(\|\Lambda_n - \Lambda\|_\infty\bigr),
\]
uniformly on $[0,\tau^\ast-\varepsilon]$. Applied with
$\Lambda_n = \widehat{\Lambda}_1$, $\Lambda = \Lambda_1$ and
$\|\widehat{\Lambda}_1 - \Lambda_1\|_\infty = O_p(n^{-1/2})$ obtained in the main text, we get
\[
\sup_{t \le \tau^\ast-\varepsilon} |\log \widehat{S}(t) + \widehat{\Lambda}_1(t)| = o_p(n^{-1/2}),
\]
which justifies the claim that KM and NA “carry the same randomness’’ at the $\sqrt{n}$ scale.

\subsection{Empirical-process expansion for SRS}
\label{supp:emp-srs}
Define
\[
\widehat{\Lambda}(t)=\int_0^t \frac{dH_{1n}(u)}{S_{Y,n}(u-)}, \qquad
\Lambda(t)=\int_0^t \frac{dH_1(u)}{S_Y(u-)}.
\]
Add and subtract $\int_0^t \frac{dH_{1n}(u)}{S_Y(u-)}$ to obtain
\[
\widehat{\Lambda}(t)-\Lambda(t)
= \underbrace{\int_0^t \frac{d(H_{1n}-H_1)(u)}{S_Y(u-)}}_{A_n(t)}
+ \underbrace{\int_0^t \Bigl(\frac{1}{S_{Y,n}(u-)} - \frac{1}{S_Y(u-)}\Bigr) dH_{1n}(u)}_{B_n(t)}.
\]
On $[0,\tau^\ast-\varepsilon]$, $H_{1n}\to H_1$ and $S_{Y,n}\to S_Y$ uniformly a.s., so
$\sup_{t \le \tau^\ast-\varepsilon} |A_n(t)| \to 0$ and
$\sup_{t \le \tau^\ast-\varepsilon} |B_n(t)| \to 0$ in probability, which yields uniform consistency of
$\widehat{\Lambda}$ and, via the product integral, of $\widehat{S}$.
Under Donsker conditions for $H_{1n}$ and $H_n$ (the indicator classes are VC),
\[
\sqrt{n}(H_{1n}-H_1) \Rightarrow \mathbb{G}_1, \qquad
\sqrt{n}(H_n-H) \Rightarrow \mathbb{G}
\]
jointly in $\ell^\infty([0,\tau^\ast-\varepsilon])$.
A first–order expansion of $x \mapsto 1/x$ gives
\[
\frac{1}{S_{Y,n}(u-)} - \frac{1}{S_Y(u-)} =
\frac{H_n(u-)-H(u-)}{S_Y(u-)^2} + o_p(n^{-1/2}),
\]
uniformly. Replacing $dH_{1n}$ by $dH_1$ in $B_n$ is $o_p(n^{-1/2})$ because $H_{1n}$ has bounded
variation and $\sup|H_n-H|=O_p(n^{-1/2})$. Consequently,
\[
\sqrt{n}(\widehat{\Lambda}-\Lambda)(t)
= \int_0^t \frac{d\mathbb{G}_1(u)}{S_Y(u-)}
+ \int_0^t \frac{\mathbb{G}(u-)}{S_Y(u-)^2} dH_1(u) + o_p(1),
\]
which is the empirical-process counterpart of the martingale CLT.

\subsection{Greenwood variance with ties (SRS)}
\label{supp:greenwood-srs}
On $[0,\tau^\ast-\varepsilon]$ we have $R(u)/n \to S_Y(u-)$ and $n^{-1}dN(u) \Rightarrow dH_1(u)$.
The NA variance is consistently estimated by
\[
\widehat{\Var}\{\widehat{\Lambda}(t)\}
= \sum_{u \le t} \frac{dN(u)}{R(u)^2},
\]
see, e.g., \textcite[Sec.~IV.1]{FlemingHarrington1991}. Applying the product–integral delta method
in Supplement~\ref{supp:prodint} with the full jump sizes $dN(u)\ge 1$ gives
\[
\widehat{\Var}\{\widehat{S}(t)\}
= \widehat{S}(t)^2 \sum_{u \le t} \frac{dN(u)}{R(u)\{R(u)-dN(u)\}},
\]
which is exactly the “with-ties’’ Greenwood estimator used in the main text and claimed to be
consistent there.

\subsection{Rank-wise i.i.d.\ structure and judgment–to–true mixing}
\label{supp:rss-iid}
\para{Rank-wise i.i.d.}

In one cycle of balanced RSS with set size $k$, we draw $k$ \emph{independent} candidate sets,
each of size $k$. In the $r$th candidate set we rank the $k$ units (by the proxy or, under perfect
ranking, by the true $X$) and \emph{only} the unit judged to be rank $r$ is measured.
Because the $k$ candidate sets are independent, the $r$-labelled measurements in different cycles are
independent and identically distributed. Repeating over $m$ cycles produces $m$ i.i.d. observations in
each rank, and different ranks use disjoint candidate sets, so they are independent across $r$.
This is the exact data layout assumed in the main CLTs.

\para{Judgment–to–true mixing}
\label{supp:mixing}

Let $T \in \{1,\dots,k\}$ denote the true rank (order statistic) and $J \in \{1,\dots,k\}$ the judged
rank under the Dell–Clutter concomitant model. For any $r$,
\[
F_r^J(t) := P(X \le t \mid J=r)
= \sum_{j=1}^k P(T=j \mid J=r) F_{[j]}(t),
\]
so $F_r^J$ is a mixture of the true rank laws.
If $P(J=r)=1/k$ for all $r$ and $P(T=j)=1/k$ for all $j$, then
\[
\frac{1}{k} \sum_{r=1}^k F_r^J(t)
= \sum_{j=1}^k \Bigl\{\frac{1}{k} \sum_{r=1}^k P(T=j \mid J=r)\Bigr\} F_{[j]}(t)
= \sum_{j=1}^k P(T=j) F_{[j]}(t)
= F(t).
\]
If $P(T=j)\neq 1/k$ the same calculation shows that the rank average converges to
$F^\star(t)=\sum_{j=1}^k P(T=j) F_{[j]}(t)$, which is what the main text notes.

\subsection{RSS empirical-process expansion and Greenwood plug-ins}
\label{supp:emp-rss}
For rank $r$, define the empirical sub-distributions based on the $m$ observations in that rank:
\[
H_{1,[r],n}(t)=\frac{1}{m}\sum_{j=1}^m I(Y_{rj} \le t, \delta_{rj}=1), \quad
H_{[r],n}(t)=\frac{1}{m}\sum_{j=1}^m I(Y_{rj} \le t), \quad
S_{Y,[r],n} = 1 - H_{[r],n}.
\]
Then
\[
\widehat{\Lambda}_{[r]}(t) = \int_0^t \frac{dH_{1,[r],n}(u)}{S_{Y,[r],n}(u-)}, \qquad
\Lambda_{[r]}(t) = \int_0^t \frac{dH_{1,[r]}(u)}{S_{Y,[r]}(u-)}.
\]
Adding and subtracting $\int_0^t \frac{dH_{1,[r],n}(u)}{S_{Y,[r]}(u-)}$ gives
\[
\widehat{\Lambda}_{[r]}(t) - \Lambda_{[r]}(t)
= \int_0^t \frac{d(H_{1,[r],n} - H_{1,[r]})(u)}{S_{Y,[r]}(u-)}
+ \int_0^t \Bigl(\frac{1}{S_{Y,[r],n}(u-)} - \frac{1}{S_{Y,[r]}(u-)}\Bigr) dH_{1,[r],n}(u),
\]
which is exactly the SRS expansion rank by rank. Since ranks are independent and
$\widehat{S}_{\mathrm{RSS}} = k^{-1} \sum_r \widehat{S}_{[r]}$, the limit covariance kernel is the
average of the $k$ within-rank kernels, as stated in the main text.

\para{Greenwood plug-ins (RSS)}
\label{supp:rss-greenwood}

For each rank $r$, the SRS Greenwood estimator
\[
\widehat{\Var}_{[r]}(t)
= \widehat{S}_{[r]}(t)^2 \sum_{u \le t} \frac{dN_r(u)}{R_r(u)\{R_r(u)-dN_r(u)\}}
\]
is consistent for the variance of $\widehat{S}_{[r]}(t)$ based on $m$ i.i.d.\ observations in that rank.
Because the RSS KM is
\[
\widehat{S}_{\mathrm{RSS}}(t) = \frac{1}{k} \sum_{r=1}^k \widehat{S}_{[r]}(t),
\]
its variance is the average of the $k$ rank variances, so the plug-in
\[
\widehat{\Var}\{\widehat{S}_{\mathrm{RSS}}(t)\}
= \frac{1}{k^2} \sum_{r=1}^k \widehat{\Var}_{[r]}(t)
\]
converges to the limit variance in the RSS CLT. The same argument holds with judged ranks
($[r]$ replaced by $J=r$).

\section*{Appendix B: AFT Simulation Tables}
\addcontentsline{toc}{section}{Appendix B: AFT Simulation Tables}



\begin{longtable}[t]{rrrrrrrrrr}
\caption{\label{tab:tab:aft_pc0_s75}t=0.35, S(t)=75\% --- AFT superpopulation: results at No censoring. Columns show KM estimates for RSS and SRS and three measures of relative efficiency (true, Monte Carlo, Greenwood). Replicates: $B_{\mathrm{MC}}=10000$, $B_{\mathrm{true}}=4000$.}\\
\toprule
$k$ & $m$ & $n$ & $\rho$ & $S(t)$ & $\widehat S_{\mathrm{RSS}}$ & $\widehat S_{\mathrm{SRS}}$ & $\mathrm{RE}_{\mathrm{true}}$ & $\mathrm{RE}_{\mathrm{MC}}$ & $\mathrm{RE}_{\mathrm{GW}}$\\
\midrule
\endfirsthead
\caption[]{t=0.35, S(t)=75\% --- AFT superpopulation: results at No censoring. Columns show KM estimates for RSS and SRS and three measures of relative efficiency (true, Monte Carlo, Greenwood). Replicates: $B_{\mathrm{MC}}=10000$, $B_{\mathrm{true}}=4000$. \textit{(continued)}}\\
\toprule
$k$ & $m$ & $n$ & $\rho$ & $S(t)$ & $\widehat S_{\mathrm{RSS}}$ & $\widehat S_{\mathrm{SRS}}$ & $\mathrm{RE}_{\mathrm{true}}$ & $\mathrm{RE}_{\mathrm{MC}}$ & $\mathrm{RE}_{\mathrm{GW}}$\\
\midrule
\endhead

\endfoot
\bottomrule
\endlastfoot
\cellcolor{gray!10}{2} & \cellcolor{gray!10}{20} & \cellcolor{gray!10}{40} & \cellcolor{gray!10}{0.1} & \cellcolor{gray!10}{0.75} & \cellcolor{gray!10}{0.750} & \cellcolor{gray!10}{0.751} & \cellcolor{gray!10}{1.033} & \cellcolor{gray!10}{1.000} & \cellcolor{gray!10}{1.028}\\
2 & 20 & 40 & 0.3 & 0.75 & 0.752 & 0.751 & 1.048 & 1.048 & 1.091\\
\cellcolor{gray!10}{2} & \cellcolor{gray!10}{20} & \cellcolor{gray!10}{40} & \cellcolor{gray!10}{0.5} & \cellcolor{gray!10}{0.75} & \cellcolor{gray!10}{0.753} & \cellcolor{gray!10}{0.752} & \cellcolor{gray!10}{1.224} & \cellcolor{gray!10}{1.223} & \cellcolor{gray!10}{1.240}\\
2 & 20 & 40 & 0.7 & 0.75 & 0.751 & 0.749 & 1.198 & 1.224 & 1.252\\
\cellcolor{gray!10}{2} & \cellcolor{gray!10}{20} & \cellcolor{gray!10}{40} & \cellcolor{gray!10}{0.9} & \cellcolor{gray!10}{0.75} & \cellcolor{gray!10}{0.749} & \cellcolor{gray!10}{0.749} & \cellcolor{gray!10}{1.315} & \cellcolor{gray!10}{1.190} & \cellcolor{gray!10}{1.243}\\
\addlinespace
2 & 50 & 100 & 0.1 & 0.75 & 0.752 & 0.751 & 0.993 & 1.004 & 1.019\\
\cellcolor{gray!10}{2} & \cellcolor{gray!10}{50} & \cellcolor{gray!10}{100} & \cellcolor{gray!10}{0.3} & \cellcolor{gray!10}{0.75} & \cellcolor{gray!10}{0.751} & \cellcolor{gray!10}{0.751} & \cellcolor{gray!10}{1.196} & \cellcolor{gray!10}{1.114} & \cellcolor{gray!10}{1.117}\\
2 & 50 & 100 & 0.5 & 0.75 & 0.750 & 0.751 & 1.178 & 1.205 & 1.221\\
\end{longtable}


\begin{longtable}[t]{rrrrrrrrrr}
\caption{\label{tab:tab:aft_pc0_s50}t=1.00, S(t)=50\% --- AFT superpopulation: results at No censoring. Columns show KM estimates for RSS and SRS and three measures of relative efficiency (true, Monte Carlo, Greenwood). Replicates: $B_{\mathrm{MC}}=10000$, $B_{\mathrm{true}}=4000$.}\\
\toprule
$k$ & $m$ & $n$ & $\rho$ & $S(t)$ & $\widehat S_{\mathrm{RSS}}$ & $\widehat S_{\mathrm{SRS}}$ & $\mathrm{RE}_{\mathrm{true}}$ & $\mathrm{RE}_{\mathrm{MC}}$ & $\mathrm{RE}_{\mathrm{GW}}$\\
\midrule
\endfirsthead
\caption[]{t=1.00, S(t)=50\% --- AFT superpopulation: results at No censoring. Columns show KM estimates for RSS and SRS and three measures of relative efficiency (true, Monte Carlo, Greenwood). Replicates: $B_{\mathrm{MC}}=10000$, $B_{\mathrm{true}}=4000$. \textit{(continued)}}\\
\toprule
$k$ & $m$ & $n$ & $\rho$ & $S(t)$ & $\widehat S_{\mathrm{RSS}}$ & $\widehat S_{\mathrm{SRS}}$ & $\mathrm{RE}_{\mathrm{true}}$ & $\mathrm{RE}_{\mathrm{MC}}$ & $\mathrm{RE}_{\mathrm{GW}}$\\
\midrule
\endhead

\endfoot
\bottomrule
\endlastfoot
\cellcolor{gray!10}{2} & \cellcolor{gray!10}{20} & \cellcolor{gray!10}{40} & \cellcolor{gray!10}{0.1} & \cellcolor{gray!10}{0.50} & \cellcolor{gray!10}{0.499} & \cellcolor{gray!10}{0.501} & \cellcolor{gray!10}{0.940} & \cellcolor{gray!10}{0.962} & \cellcolor{gray!10}{1.035}\\
2 & 20 & 40 & 0.3 & 0.50 & 0.500 & 0.499 & 1.076 & 1.075 & 1.104\\
\cellcolor{gray!10}{2} & \cellcolor{gray!10}{20} & \cellcolor{gray!10}{40} & \cellcolor{gray!10}{0.5} & \cellcolor{gray!10}{0.50} & \cellcolor{gray!10}{0.501} & \cellcolor{gray!10}{0.501} & \cellcolor{gray!10}{1.258} & \cellcolor{gray!10}{1.282} & \cellcolor{gray!10}{1.325}\\
2 & 20 & 40 & 0.7 & 0.50 & 0.500 & 0.499 & 1.290 & 1.295 & 1.335\\
\cellcolor{gray!10}{2} & \cellcolor{gray!10}{20} & \cellcolor{gray!10}{40} & \cellcolor{gray!10}{0.9} & \cellcolor{gray!10}{0.50} & \cellcolor{gray!10}{0.498} & \cellcolor{gray!10}{0.498} & \cellcolor{gray!10}{1.335} & \cellcolor{gray!10}{1.313} & \cellcolor{gray!10}{1.331}\\
\addlinespace
2 & 50 & 100 & 0.1 & 0.50 & 0.500 & 0.500 & 0.990 & 1.017 & 1.017\\
\cellcolor{gray!10}{2} & \cellcolor{gray!10}{50} & \cellcolor{gray!10}{100} & \cellcolor{gray!10}{0.3} & \cellcolor{gray!10}{0.50} & \cellcolor{gray!10}{0.501} & \cellcolor{gray!10}{0.500} & \cellcolor{gray!10}{1.185} & \cellcolor{gray!10}{1.131} & \cellcolor{gray!10}{1.147}\\
2 & 50 & 100 & 0.5 & 0.50 & 0.500 & 0.501 & 1.274 & 1.232 & 1.312\\
\end{longtable}


\begin{longtable}[t]{rrrrrrrrrr}
\caption{\label{tab:tab:aft_pc0_s25}t=2.85, S(t)=25\% --- AFT superpopulation: results at No censoring. Columns show KM estimates for RSS and SRS and three measures of relative efficiency (true, Monte Carlo, Greenwood). Replicates: $B_{\mathrm{MC}}=10000$, $B_{\mathrm{true}}=4000$.}\\
\toprule
$k$ & $m$ & $n$ & $\rho$ & $S(t)$ & $\widehat S_{\mathrm{RSS}}$ & $\widehat S_{\mathrm{SRS}}$ & $\mathrm{RE}_{\mathrm{true}}$ & $\mathrm{RE}_{\mathrm{MC}}$ & $\mathrm{RE}_{\mathrm{GW}}$\\
\midrule
\endfirsthead
\caption[]{t=2.85, S(t)=25\% --- AFT superpopulation: results at No censoring. Columns show KM estimates for RSS and SRS and three measures of relative efficiency (true, Monte Carlo, Greenwood). Replicates: $B_{\mathrm{MC}}=10000$, $B_{\mathrm{true}}=4000$. \textit{(continued)}}\\
\toprule
$k$ & $m$ & $n$ & $\rho$ & $S(t)$ & $\widehat S_{\mathrm{RSS}}$ & $\widehat S_{\mathrm{SRS}}$ & $\mathrm{RE}_{\mathrm{true}}$ & $\mathrm{RE}_{\mathrm{MC}}$ & $\mathrm{RE}_{\mathrm{GW}}$\\
\midrule
\endhead

\endfoot
\bottomrule
\endlastfoot
\cellcolor{gray!10}{2} & \cellcolor{gray!10}{20} & \cellcolor{gray!10}{40} & \cellcolor{gray!10}{0.1} & \cellcolor{gray!10}{0.25} & \cellcolor{gray!10}{0.250} & \cellcolor{gray!10}{0.253} & \cellcolor{gray!10}{0.980} & \cellcolor{gray!10}{0.988} & \cellcolor{gray!10}{1.040}\\
2 & 20 & 40 & 0.3 & 0.25 & 0.253 & 0.250 & 1.137 & 1.094 & 1.079\\
\cellcolor{gray!10}{2} & \cellcolor{gray!10}{20} & \cellcolor{gray!10}{40} & \cellcolor{gray!10}{0.5} & \cellcolor{gray!10}{0.25} & \cellcolor{gray!10}{0.255} & \cellcolor{gray!10}{0.250} & \cellcolor{gray!10}{1.292} & \cellcolor{gray!10}{1.319} & \cellcolor{gray!10}{1.197}\\
2 & 20 & 40 & 0.7 & 0.25 & 0.256 & 0.249 & 1.240 & 1.285 & 1.196\\
\cellcolor{gray!10}{2} & \cellcolor{gray!10}{20} & \cellcolor{gray!10}{40} & \cellcolor{gray!10}{0.9} & \cellcolor{gray!10}{0.25} & \cellcolor{gray!10}{0.256} & \cellcolor{gray!10}{0.249} & \cellcolor{gray!10}{1.306} & \cellcolor{gray!10}{1.318} & \cellcolor{gray!10}{1.193}\\
\addlinespace
2 & 50 & 100 & 0.1 & 0.25 & 0.250 & 0.250 & 0.980 & 0.982 & 1.017\\
\cellcolor{gray!10}{2} & \cellcolor{gray!10}{50} & \cellcolor{gray!10}{100} & \cellcolor{gray!10}{0.3} & \cellcolor{gray!10}{0.25} & \cellcolor{gray!10}{0.250} & \cellcolor{gray!10}{0.249} & \cellcolor{gray!10}{1.099} & \cellcolor{gray!10}{1.118} & \cellcolor{gray!10}{1.115}\\
2 & 50 & 100 & 0.5 & 0.25 & 0.251 & 0.250 & 1.249 & 1.234 & 1.222\\
\end{longtable}


\begin{longtable}[t]{rrrrrrrrrr}
\caption{\label{tab:tab:aft_pc0_s10}t=7.30, S(t)=10\% --- AFT superpopulation: results at No censoring. Columns show KM estimates for RSS and SRS and three measures of relative efficiency (true, Monte Carlo, Greenwood). Replicates: $B_{\mathrm{MC}}=10000$, $B_{\mathrm{true}}=4000$.}\\
\toprule
$k$ & $m$ & $n$ & $\rho$ & $S(t)$ & $\widehat S_{\mathrm{RSS}}$ & $\widehat S_{\mathrm{SRS}}$ & $\mathrm{RE}_{\mathrm{true}}$ & $\mathrm{RE}_{\mathrm{MC}}$ & $\mathrm{RE}_{\mathrm{GW}}$\\
\midrule
\endfirsthead
\caption[]{t=7.30, S(t)=10\% --- AFT superpopulation: results at No censoring. Columns show KM estimates for RSS and SRS and three measures of relative efficiency (true, Monte Carlo, Greenwood). Replicates: $B_{\mathrm{MC}}=10000$, $B_{\mathrm{true}}=4000$. \textit{(continued)}}\\
\toprule
$k$ & $m$ & $n$ & $\rho$ & $S(t)$ & $\widehat S_{\mathrm{RSS}}$ & $\widehat S_{\mathrm{SRS}}$ & $\mathrm{RE}_{\mathrm{true}}$ & $\mathrm{RE}_{\mathrm{MC}}$ & $\mathrm{RE}_{\mathrm{GW}}$\\
\midrule
\endhead

\endfoot
\bottomrule
\endlastfoot
\cellcolor{gray!10}{2} & \cellcolor{gray!10}{20} & \cellcolor{gray!10}{40} & \cellcolor{gray!10}{0.1} & \cellcolor{gray!10}{0.10} & \cellcolor{gray!10}{0.107} & \cellcolor{gray!10}{0.102} & \cellcolor{gray!10}{1.250} & \cellcolor{gray!10}{1.252} & \cellcolor{gray!10}{0.975}\\
2 & 20 & 40 & 0.3 & 0.10 & 0.112 & 0.101 & 1.211 & 1.250 & 0.944\\
\cellcolor{gray!10}{2} & \cellcolor{gray!10}{20} & \cellcolor{gray!10}{40} & \cellcolor{gray!10}{0.5} & \cellcolor{gray!10}{0.10} & \cellcolor{gray!10}{0.118} & \cellcolor{gray!10}{0.100} & \cellcolor{gray!10}{1.185} & \cellcolor{gray!10}{1.172} & \cellcolor{gray!10}{0.911}\\
2 & 20 & 40 & 0.7 & 0.10 & 0.120 & 0.100 & 1.143 & 1.153 & 0.902\\
\cellcolor{gray!10}{2} & \cellcolor{gray!10}{20} & \cellcolor{gray!10}{40} & \cellcolor{gray!10}{0.9} & \cellcolor{gray!10}{0.10} & \cellcolor{gray!10}{0.121} & \cellcolor{gray!10}{0.100} & \cellcolor{gray!10}{1.173} & \cellcolor{gray!10}{1.158} & \cellcolor{gray!10}{0.897}\\
\addlinespace
2 & 50 & 100 & 0.1 & 0.10 & 0.100 & 0.100 & 1.050 & 1.014 & 1.014\\
\cellcolor{gray!10}{2} & \cellcolor{gray!10}{50} & \cellcolor{gray!10}{100} & \cellcolor{gray!10}{0.3} & \cellcolor{gray!10}{0.10} & \cellcolor{gray!10}{0.103} & \cellcolor{gray!10}{0.100} & \cellcolor{gray!10}{1.109} & \cellcolor{gray!10}{1.108} & \cellcolor{gray!10}{1.038}\\
2 & 50 & 100 & 0.5 & 0.10 & 0.105 & 0.100 & 1.243 & 1.163 & 1.036\\
\end{longtable}


\begin{longtable}[t]{rrrrrrrrrr}
\caption{\label{tab:tab:aft_pc10_s75}t=0.35, S(t)=75\% --- AFT superpopulation: results at 10\% censoring. Columns show KM estimates for RSS and SRS and three measures of relative efficiency (true, Monte Carlo, Greenwood). Replicates: $B_{\mathrm{MC}}=10000$, $B_{\mathrm{true}}=4000$.}\\
\toprule
$k$ & $m$ & $n$ & $\rho$ & $S(t)$ & $\widehat S_{\mathrm{RSS}}$ & $\widehat S_{\mathrm{SRS}}$ & $\mathrm{RE}_{\mathrm{true}}$ & $\mathrm{RE}_{\mathrm{MC}}$ & $\mathrm{RE}_{\mathrm{GW}}$\\
\midrule
\endfirsthead
\caption[]{t=0.35, S(t)=75\% --- AFT superpopulation: results at 10\% censoring. Columns show KM estimates for RSS and SRS and three measures of relative efficiency (true, Monte Carlo, Greenwood). Replicates: $B_{\mathrm{MC}}=10000$, $B_{\mathrm{true}}=4000$. \textit{(continued)}}\\
\toprule
$k$ & $m$ & $n$ & $\rho$ & $S(t)$ & $\widehat S_{\mathrm{RSS}}$ & $\widehat S_{\mathrm{SRS}}$ & $\mathrm{RE}_{\mathrm{true}}$ & $\mathrm{RE}_{\mathrm{MC}}$ & $\mathrm{RE}_{\mathrm{GW}}$\\
\midrule
\endhead

\endfoot
\bottomrule
\endlastfoot
\cellcolor{gray!10}{2} & \cellcolor{gray!10}{20} & \cellcolor{gray!10}{40} & \cellcolor{gray!10}{0.1} & \cellcolor{gray!10}{0.75} & \cellcolor{gray!10}{0.747} & \cellcolor{gray!10}{0.750} & \cellcolor{gray!10}{1.016} & \cellcolor{gray!10}{0.977} & \cellcolor{gray!10}{1.026}\\
2 & 20 & 40 & 0.3 & 0.75 & 0.749 & 0.749 & 1.013 & 1.097 & 1.090\\
\cellcolor{gray!10}{2} & \cellcolor{gray!10}{20} & \cellcolor{gray!10}{40} & \cellcolor{gray!10}{0.5} & \cellcolor{gray!10}{0.75} & \cellcolor{gray!10}{0.750} & \cellcolor{gray!10}{0.750} & \cellcolor{gray!10}{1.226} & \cellcolor{gray!10}{1.179} & \cellcolor{gray!10}{1.233}\\
2 & 20 & 40 & 0.7 & 0.75 & 0.751 & 0.750 & 1.207 & 1.235 & 1.242\\
\cellcolor{gray!10}{2} & \cellcolor{gray!10}{20} & \cellcolor{gray!10}{40} & \cellcolor{gray!10}{0.9} & \cellcolor{gray!10}{0.75} & \cellcolor{gray!10}{0.751} & \cellcolor{gray!10}{0.749} & \cellcolor{gray!10}{1.221} & \cellcolor{gray!10}{1.222} & \cellcolor{gray!10}{1.245}\\
\addlinespace
2 & 50 & 100 & 0.1 & 0.75 & 0.749 & 0.749 & 1.024 & 0.995 & 1.017\\
\cellcolor{gray!10}{2} & \cellcolor{gray!10}{50} & \cellcolor{gray!10}{100} & \cellcolor{gray!10}{0.3} & \cellcolor{gray!10}{0.75} & \cellcolor{gray!10}{0.749} & \cellcolor{gray!10}{0.749} & \cellcolor{gray!10}{1.049} & \cellcolor{gray!10}{1.037} & \cellcolor{gray!10}{1.071}\\
2 & 50 & 100 & 0.5 & 0.75 & 0.751 & 0.751 & 1.157 & 1.221 & 1.202\\
\end{longtable}


\begin{longtable}[t]{rrrrrrrrrr}
\caption{\label{tab:tab:aft_pc10_s50}t=1.00, S(t)=50\% --- AFT superpopulation: results at 10\% censoring. Columns show KM estimates for RSS and SRS and three measures of relative efficiency (true, Monte Carlo, Greenwood). Replicates: $B_{\mathrm{MC}}=10000$, $B_{\mathrm{true}}=4000$.}\\
\toprule
$k$ & $m$ & $n$ & $\rho$ & $S(t)$ & $\widehat S_{\mathrm{RSS}}$ & $\widehat S_{\mathrm{SRS}}$ & $\mathrm{RE}_{\mathrm{true}}$ & $\mathrm{RE}_{\mathrm{MC}}$ & $\mathrm{RE}_{\mathrm{GW}}$\\
\midrule
\endfirsthead
\caption[]{t=1.00, S(t)=50\% --- AFT superpopulation: results at 10\% censoring. Columns show KM estimates for RSS and SRS and three measures of relative efficiency (true, Monte Carlo, Greenwood). Replicates: $B_{\mathrm{MC}}=10000$, $B_{\mathrm{true}}=4000$. \textit{(continued)}}\\
\toprule
$k$ & $m$ & $n$ & $\rho$ & $S(t)$ & $\widehat S_{\mathrm{RSS}}$ & $\widehat S_{\mathrm{SRS}}$ & $\mathrm{RE}_{\mathrm{true}}$ & $\mathrm{RE}_{\mathrm{MC}}$ & $\mathrm{RE}_{\mathrm{GW}}$\\
\midrule
\endhead

\endfoot
\bottomrule
\endlastfoot
\cellcolor{gray!10}{2} & \cellcolor{gray!10}{20} & \cellcolor{gray!10}{40} & \cellcolor{gray!10}{0.1} & \cellcolor{gray!10}{0.50} & \cellcolor{gray!10}{0.499} & \cellcolor{gray!10}{0.501} & \cellcolor{gray!10}{1.040} & \cellcolor{gray!10}{1.017} & \cellcolor{gray!10}{1.034}\\
2 & 20 & 40 & 0.3 & 0.50 & 0.500 & 0.502 & 1.021 & 1.102 & 1.103\\
\cellcolor{gray!10}{2} & \cellcolor{gray!10}{20} & \cellcolor{gray!10}{40} & \cellcolor{gray!10}{0.5} & \cellcolor{gray!10}{0.50} & \cellcolor{gray!10}{0.499} & \cellcolor{gray!10}{0.501} & \cellcolor{gray!10}{1.274} & \cellcolor{gray!10}{1.278} & \cellcolor{gray!10}{1.311}\\
2 & 20 & 40 & 0.7 & 0.50 & 0.501 & 0.500 & 1.259 & 1.304 & 1.324\\
\cellcolor{gray!10}{2} & \cellcolor{gray!10}{20} & \cellcolor{gray!10}{40} & \cellcolor{gray!10}{0.9} & \cellcolor{gray!10}{0.50} & \cellcolor{gray!10}{0.502} & \cellcolor{gray!10}{0.501} & \cellcolor{gray!10}{1.291} & \cellcolor{gray!10}{1.326} & \cellcolor{gray!10}{1.324}\\
\addlinespace
2 & 50 & 100 & 0.1 & 0.50 & 0.500 & 0.500 & 1.024 & 1.035 & 1.017\\
\cellcolor{gray!10}{2} & \cellcolor{gray!10}{50} & \cellcolor{gray!10}{100} & \cellcolor{gray!10}{0.3} & \cellcolor{gray!10}{0.50} & \cellcolor{gray!10}{0.499} & \cellcolor{gray!10}{0.499} & \cellcolor{gray!10}{1.113} & \cellcolor{gray!10}{1.071} & \cellcolor{gray!10}{1.087}\\
2 & 50 & 100 & 0.5 & 0.50 & 0.502 & 0.501 & 1.241 & 1.242 & 1.271\\
\end{longtable}


\begin{longtable}[t]{rrrrrrrrrr}
\caption{\label{tab:tab:aft_pc10_s25}t=2.85, S(t)=25\% --- AFT superpopulation: results at 10\% censoring. Columns show KM estimates for RSS and SRS and three measures of relative efficiency (true, Monte Carlo, Greenwood). Replicates: $B_{\mathrm{MC}}=10000$, $B_{\mathrm{true}}=4000$.}\\
\toprule
$k$ & $m$ & $n$ & $\rho$ & $S(t)$ & $\widehat S_{\mathrm{RSS}}$ & $\widehat S_{\mathrm{SRS}}$ & $\mathrm{RE}_{\mathrm{true}}$ & $\mathrm{RE}_{\mathrm{MC}}$ & $\mathrm{RE}_{\mathrm{GW}}$\\
\midrule
\endfirsthead
\caption[]{t=2.85, S(t)=25\% --- AFT superpopulation: results at 10\% censoring. Columns show KM estimates for RSS and SRS and three measures of relative efficiency (true, Monte Carlo, Greenwood). Replicates: $B_{\mathrm{MC}}=10000$, $B_{\mathrm{true}}=4000$. \textit{(continued)}}\\
\toprule
$k$ & $m$ & $n$ & $\rho$ & $S(t)$ & $\widehat S_{\mathrm{RSS}}$ & $\widehat S_{\mathrm{SRS}}$ & $\mathrm{RE}_{\mathrm{true}}$ & $\mathrm{RE}_{\mathrm{MC}}$ & $\mathrm{RE}_{\mathrm{GW}}$\\
\midrule
\endhead

\endfoot
\bottomrule
\endlastfoot
\cellcolor{gray!10}{2} & \cellcolor{gray!10}{20} & \cellcolor{gray!10}{40} & \cellcolor{gray!10}{0.1} & \cellcolor{gray!10}{0.25} & \cellcolor{gray!10}{0.250} & \cellcolor{gray!10}{0.250} & \cellcolor{gray!10}{1.069} & \cellcolor{gray!10}{1.009} & \cellcolor{gray!10}{1.033}\\
2 & 20 & 40 & 0.3 & 0.25 & 0.252 & 0.250 & 1.056 & 1.125 & 1.080\\
\cellcolor{gray!10}{2} & \cellcolor{gray!10}{20} & \cellcolor{gray!10}{40} & \cellcolor{gray!10}{0.5} & \cellcolor{gray!10}{0.25} & \cellcolor{gray!10}{0.256} & \cellcolor{gray!10}{0.250} & \cellcolor{gray!10}{1.304} & \cellcolor{gray!10}{1.323} & \cellcolor{gray!10}{1.181}\\
2 & 20 & 40 & 0.7 & 0.25 & 0.258 & 0.251 & 1.280 & 1.270 & 1.184\\
\cellcolor{gray!10}{2} & \cellcolor{gray!10}{20} & \cellcolor{gray!10}{40} & \cellcolor{gray!10}{0.9} & \cellcolor{gray!10}{0.25} & \cellcolor{gray!10}{0.258} & \cellcolor{gray!10}{0.251} & \cellcolor{gray!10}{1.339} & \cellcolor{gray!10}{1.332} & \cellcolor{gray!10}{1.183}\\
\addlinespace
2 & 50 & 100 & 0.1 & 0.25 & 0.249 & 0.250 & 1.027 & 0.992 & 1.018\\
\cellcolor{gray!10}{2} & \cellcolor{gray!10}{50} & \cellcolor{gray!10}{100} & \cellcolor{gray!10}{0.3} & \cellcolor{gray!10}{0.25} & \cellcolor{gray!10}{0.249} & \cellcolor{gray!10}{0.249} & \cellcolor{gray!10}{1.012} & \cellcolor{gray!10}{1.076} & \cellcolor{gray!10}{1.071}\\
2 & 50 & 100 & 0.5 & 0.25 & 0.251 & 0.250 & 1.258 & 1.170 & 1.193\\
\end{longtable}


\begin{longtable}[t]{rrrrrrrrrr}
\caption{\label{tab:tab:aft_pc10_s10}t=7.30, S(t)=10\% --- AFT superpopulation: results at 10\% censoring. Columns show KM estimates for RSS and SRS and three measures of relative efficiency (true, Monte Carlo, Greenwood). Replicates: $B_{\mathrm{MC}}=10000$, $B_{\mathrm{true}}=4000$.}\\
\toprule
$k$ & $m$ & $n$ & $\rho$ & $S(t)$ & $\widehat S_{\mathrm{RSS}}$ & $\widehat S_{\mathrm{SRS}}$ & $\mathrm{RE}_{\mathrm{true}}$ & $\mathrm{RE}_{\mathrm{MC}}$ & $\mathrm{RE}_{\mathrm{GW}}$\\
\midrule
\endfirsthead
\caption[]{t=7.30, S(t)=10\% --- AFT superpopulation: results at 10\% censoring. Columns show KM estimates for RSS and SRS and three measures of relative efficiency (true, Monte Carlo, Greenwood). Replicates: $B_{\mathrm{MC}}=10000$, $B_{\mathrm{true}}=4000$. \textit{(continued)}}\\
\toprule
$k$ & $m$ & $n$ & $\rho$ & $S(t)$ & $\widehat S_{\mathrm{RSS}}$ & $\widehat S_{\mathrm{SRS}}$ & $\mathrm{RE}_{\mathrm{true}}$ & $\mathrm{RE}_{\mathrm{MC}}$ & $\mathrm{RE}_{\mathrm{GW}}$\\
\midrule
\endhead

\endfoot
\bottomrule
\endlastfoot
\cellcolor{gray!10}{2} & \cellcolor{gray!10}{20} & \cellcolor{gray!10}{40} & \cellcolor{gray!10}{0.1} & \cellcolor{gray!10}{0.10} & \cellcolor{gray!10}{0.112} & \cellcolor{gray!10}{0.101} & \cellcolor{gray!10}{1.347} & \cellcolor{gray!10}{1.322} & \cellcolor{gray!10}{0.937}\\
2 & 20 & 40 & 0.3 & 0.10 & 0.116 & 0.102 & 1.325 & 1.293 & 0.931\\
\cellcolor{gray!10}{2} & \cellcolor{gray!10}{20} & \cellcolor{gray!10}{40} & \cellcolor{gray!10}{0.5} & \cellcolor{gray!10}{0.10} & \cellcolor{gray!10}{0.123} & \cellcolor{gray!10}{0.102} & \cellcolor{gray!10}{1.192} & \cellcolor{gray!10}{1.178} & \cellcolor{gray!10}{0.901}\\
2 & 20 & 40 & 0.7 & 0.10 & 0.125 & 0.102 & 1.155 & 1.135 & 0.895\\
\cellcolor{gray!10}{2} & \cellcolor{gray!10}{20} & \cellcolor{gray!10}{40} & \cellcolor{gray!10}{0.9} & \cellcolor{gray!10}{0.10} & \cellcolor{gray!10}{0.123} & \cellcolor{gray!10}{0.101} & \cellcolor{gray!10}{1.147} & \cellcolor{gray!10}{1.181} & \cellcolor{gray!10}{0.899}\\
\addlinespace
2 & 50 & 100 & 0.1 & 0.10 & 0.100 & 0.100 & 1.001 & 1.053 & 1.013\\
\cellcolor{gray!10}{2} & \cellcolor{gray!10}{50} & \cellcolor{gray!10}{100} & \cellcolor{gray!10}{0.3} & \cellcolor{gray!10}{0.10} & \cellcolor{gray!10}{0.102} & \cellcolor{gray!10}{0.100} & \cellcolor{gray!10}{1.049} & \cellcolor{gray!10}{1.119} & \cellcolor{gray!10}{1.018}\\
2 & 50 & 100 & 0.5 & 0.10 & 0.106 & 0.100 & 1.177 & 1.145 & 1.019\\
\end{longtable}


\begin{longtable}[t]{rrrrrrrrrr}
\caption{\label{tab:tab:aft_pc30_s75}t=0.35, S(t)=75\% --- AFT superpopulation: results at 30\% censoring. Columns show KM estimates for RSS and SRS and three measures of relative efficiency (true, Monte Carlo, Greenwood). Replicates: $B_{\mathrm{MC}}=10000$, $B_{\mathrm{true}}=4000$.}\\
\toprule
$k$ & $m$ & $n$ & $\rho$ & $S(t)$ & $\widehat S_{\mathrm{RSS}}$ & $\widehat S_{\mathrm{SRS}}$ & $\mathrm{RE}_{\mathrm{true}}$ & $\mathrm{RE}_{\mathrm{MC}}$ & $\mathrm{RE}_{\mathrm{GW}}$\\
\midrule
\endfirsthead
\caption[]{t=0.35, S(t)=75\% --- AFT superpopulation: results at 30\% censoring. Columns show KM estimates for RSS and SRS and three measures of relative efficiency (true, Monte Carlo, Greenwood). Replicates: $B_{\mathrm{MC}}=10000$, $B_{\mathrm{true}}=4000$. \textit{(continued)}}\\
\toprule
$k$ & $m$ & $n$ & $\rho$ & $S(t)$ & $\widehat S_{\mathrm{RSS}}$ & $\widehat S_{\mathrm{SRS}}$ & $\mathrm{RE}_{\mathrm{true}}$ & $\mathrm{RE}_{\mathrm{MC}}$ & $\mathrm{RE}_{\mathrm{GW}}$\\
\midrule
\endhead

\endfoot
\bottomrule
\endlastfoot
\cellcolor{gray!10}{2} & \cellcolor{gray!10}{20} & \cellcolor{gray!10}{40} & \cellcolor{gray!10}{0.1} & \cellcolor{gray!10}{0.75} & \cellcolor{gray!10}{0.750} & \cellcolor{gray!10}{0.752} & \cellcolor{gray!10}{0.987} & \cellcolor{gray!10}{0.965} & \cellcolor{gray!10}{1.032}\\
2 & 20 & 40 & 0.3 & 0.75 & 0.750 & 0.750 & 1.038 & 1.031 & 1.089\\
\cellcolor{gray!10}{2} & \cellcolor{gray!10}{20} & \cellcolor{gray!10}{40} & \cellcolor{gray!10}{0.5} & \cellcolor{gray!10}{0.75} & \cellcolor{gray!10}{0.749} & \cellcolor{gray!10}{0.749} & \cellcolor{gray!10}{1.176} & \cellcolor{gray!10}{1.242} & \cellcolor{gray!10}{1.243}\\
2 & 20 & 40 & 0.7 & 0.75 & 0.750 & 0.750 & 1.250 & 1.196 & 1.240\\
\cellcolor{gray!10}{2} & \cellcolor{gray!10}{20} & \cellcolor{gray!10}{40} & \cellcolor{gray!10}{0.9} & \cellcolor{gray!10}{0.75} & \cellcolor{gray!10}{0.750} & \cellcolor{gray!10}{0.750} & \cellcolor{gray!10}{1.148} & \cellcolor{gray!10}{1.216} & \cellcolor{gray!10}{1.239}\\
\addlinespace
2 & 50 & 100 & 0.1 & 0.75 & 0.751 & 0.751 & 1.063 & 1.018 & 1.022\\
\cellcolor{gray!10}{2} & \cellcolor{gray!10}{50} & \cellcolor{gray!10}{100} & \cellcolor{gray!10}{0.3} & \cellcolor{gray!10}{0.75} & \cellcolor{gray!10}{0.752} & \cellcolor{gray!10}{0.752} & \cellcolor{gray!10}{1.030} & \cellcolor{gray!10}{1.068} & \cellcolor{gray!10}{1.071}\\
2 & 50 & 100 & 0.5 & 0.75 & 0.750 & 0.750 & 1.153 & 1.203 & 1.224\\
\end{longtable}


\begin{longtable}[t]{rrrrrrrrrr}
\caption{\label{tab:tab:aft_pc30_s50}t=1.00, S(t)=50\% --- AFT superpopulation: results at 30\% censoring. Columns show KM estimates for RSS and SRS and three measures of relative efficiency (true, Monte Carlo, Greenwood). Replicates: $B_{\mathrm{MC}}=10000$, $B_{\mathrm{true}}=4000$.}\\
\toprule
$k$ & $m$ & $n$ & $\rho$ & $S(t)$ & $\widehat S_{\mathrm{RSS}}$ & $\widehat S_{\mathrm{SRS}}$ & $\mathrm{RE}_{\mathrm{true}}$ & $\mathrm{RE}_{\mathrm{MC}}$ & $\mathrm{RE}_{\mathrm{GW}}$\\
\midrule
\endfirsthead
\caption[]{t=1.00, S(t)=50\% --- AFT superpopulation: results at 30\% censoring. Columns show KM estimates for RSS and SRS and three measures of relative efficiency (true, Monte Carlo, Greenwood). Replicates: $B_{\mathrm{MC}}=10000$, $B_{\mathrm{true}}=4000$. \textit{(continued)}}\\
\toprule
$k$ & $m$ & $n$ & $\rho$ & $S(t)$ & $\widehat S_{\mathrm{RSS}}$ & $\widehat S_{\mathrm{SRS}}$ & $\mathrm{RE}_{\mathrm{true}}$ & $\mathrm{RE}_{\mathrm{MC}}$ & $\mathrm{RE}_{\mathrm{GW}}$\\
\midrule
\endhead

\endfoot
\bottomrule
\endlastfoot
\cellcolor{gray!10}{2} & \cellcolor{gray!10}{20} & \cellcolor{gray!10}{40} & \cellcolor{gray!10}{0.1} & \cellcolor{gray!10}{0.50} & \cellcolor{gray!10}{0.499} & \cellcolor{gray!10}{0.500} & \cellcolor{gray!10}{0.983} & \cellcolor{gray!10}{1.007} & \cellcolor{gray!10}{1.040}\\
2 & 20 & 40 & 0.3 & 0.50 & 0.499 & 0.500 & 1.044 & 1.076 & 1.104\\
\cellcolor{gray!10}{2} & \cellcolor{gray!10}{20} & \cellcolor{gray!10}{40} & \cellcolor{gray!10}{0.5} & \cellcolor{gray!10}{0.50} & \cellcolor{gray!10}{0.500} & \cellcolor{gray!10}{0.501} & \cellcolor{gray!10}{1.276} & \cellcolor{gray!10}{1.304} & \cellcolor{gray!10}{1.322}\\
2 & 20 & 40 & 0.7 & 0.50 & 0.501 & 0.501 & 1.276 & 1.338 & 1.322\\
\cellcolor{gray!10}{2} & \cellcolor{gray!10}{20} & \cellcolor{gray!10}{40} & \cellcolor{gray!10}{0.9} & \cellcolor{gray!10}{0.50} & \cellcolor{gray!10}{0.502} & \cellcolor{gray!10}{0.500} & \cellcolor{gray!10}{1.200} & \cellcolor{gray!10}{1.276} & \cellcolor{gray!10}{1.319}\\
\addlinespace
2 & 50 & 100 & 0.1 & 0.50 & 0.501 & 0.501 & 1.058 & 1.000 & 1.021\\
\cellcolor{gray!10}{2} & \cellcolor{gray!10}{50} & \cellcolor{gray!10}{100} & \cellcolor{gray!10}{0.3} & \cellcolor{gray!10}{0.50} & \cellcolor{gray!10}{0.501} & \cellcolor{gray!10}{0.501} & \cellcolor{gray!10}{1.079} & \cellcolor{gray!10}{1.086} & \cellcolor{gray!10}{1.084}\\
2 & 50 & 100 & 0.5 & 0.50 & 0.500 & 0.500 & 1.296 & 1.293 & 1.302\\
\end{longtable}


\begin{longtable}[t]{rrrrrrrrrr}
\caption{\label{tab:tab:aft_pc30_s25}t=2.85, S(t)=25\% --- AFT superpopulation: results at 30\% censoring. Columns show KM estimates for RSS and SRS and three measures of relative efficiency (true, Monte Carlo, Greenwood). Replicates: $B_{\mathrm{MC}}=10000$, $B_{\mathrm{true}}=4000$.}\\
\toprule
$k$ & $m$ & $n$ & $\rho$ & $S(t)$ & $\widehat S_{\mathrm{RSS}}$ & $\widehat S_{\mathrm{SRS}}$ & $\mathrm{RE}_{\mathrm{true}}$ & $\mathrm{RE}_{\mathrm{MC}}$ & $\mathrm{RE}_{\mathrm{GW}}$\\
\midrule
\endfirsthead
\caption[]{t=2.85, S(t)=25\% --- AFT superpopulation: results at 30\% censoring. Columns show KM estimates for RSS and SRS and three measures of relative efficiency (true, Monte Carlo, Greenwood). Replicates: $B_{\mathrm{MC}}=10000$, $B_{\mathrm{true}}=4000$. \textit{(continued)}}\\
\toprule
$k$ & $m$ & $n$ & $\rho$ & $S(t)$ & $\widehat S_{\mathrm{RSS}}$ & $\widehat S_{\mathrm{SRS}}$ & $\mathrm{RE}_{\mathrm{true}}$ & $\mathrm{RE}_{\mathrm{MC}}$ & $\mathrm{RE}_{\mathrm{GW}}$\\
\midrule
\endhead

\endfoot
\bottomrule
\endlastfoot
\cellcolor{gray!10}{2} & \cellcolor{gray!10}{20} & \cellcolor{gray!10}{40} & \cellcolor{gray!10}{0.1} & \cellcolor{gray!10}{0.25} & \cellcolor{gray!10}{0.251} & \cellcolor{gray!10}{0.251} & \cellcolor{gray!10}{1.034} & \cellcolor{gray!10}{1.061} & \cellcolor{gray!10}{1.038}\\
2 & 20 & 40 & 0.3 & 0.25 & 0.253 & 0.250 & 1.168 & 1.137 & 1.075\\
\cellcolor{gray!10}{2} & \cellcolor{gray!10}{20} & \cellcolor{gray!10}{40} & \cellcolor{gray!10}{0.5} & \cellcolor{gray!10}{0.25} & \cellcolor{gray!10}{0.261} & \cellcolor{gray!10}{0.249} & \cellcolor{gray!10}{1.337} & \cellcolor{gray!10}{1.312} & \cellcolor{gray!10}{1.154}\\
2 & 20 & 40 & 0.7 & 0.25 & 0.262 & 0.250 & 1.367 & 1.269 & 1.155\\
\cellcolor{gray!10}{2} & \cellcolor{gray!10}{20} & \cellcolor{gray!10}{40} & \cellcolor{gray!10}{0.9} & \cellcolor{gray!10}{0.25} & \cellcolor{gray!10}{0.261} & \cellcolor{gray!10}{0.250} & \cellcolor{gray!10}{1.278} & \cellcolor{gray!10}{1.301} & \cellcolor{gray!10}{1.154}\\
\addlinespace
2 & 50 & 100 & 0.1 & 0.25 & 0.250 & 0.250 & 1.010 & 0.991 & 1.020\\
\cellcolor{gray!10}{2} & \cellcolor{gray!10}{50} & \cellcolor{gray!10}{100} & \cellcolor{gray!10}{0.3} & \cellcolor{gray!10}{0.25} & \cellcolor{gray!10}{0.251} & \cellcolor{gray!10}{0.251} & \cellcolor{gray!10}{1.053} & \cellcolor{gray!10}{1.086} & \cellcolor{gray!10}{1.067}\\
2 & 50 & 100 & 0.5 & 0.25 & 0.252 & 0.250 & 1.202 & 1.204 & 1.197\\
\end{longtable}


\begin{longtable}[t]{rrrrrrrrrr}
\caption{\label{tab:tab:aft_pc30_s10}t=7.30, S(t)=10\% --- AFT superpopulation: results at 30\% censoring. Columns show KM estimates for RSS and SRS and three measures of relative efficiency (true, Monte Carlo, Greenwood). Replicates: $B_{\mathrm{MC}}=10000$, $B_{\mathrm{true}}=4000$.}\\
\toprule
$k$ & $m$ & $n$ & $\rho$ & $S(t)$ & $\widehat S_{\mathrm{RSS}}$ & $\widehat S_{\mathrm{SRS}}$ & $\mathrm{RE}_{\mathrm{true}}$ & $\mathrm{RE}_{\mathrm{MC}}$ & $\mathrm{RE}_{\mathrm{GW}}$\\
\midrule
\endfirsthead
\caption[]{t=7.30, S(t)=10\% --- AFT superpopulation: results at 30\% censoring. Columns show KM estimates for RSS and SRS and three measures of relative efficiency (true, Monte Carlo, Greenwood). Replicates: $B_{\mathrm{MC}}=10000$, $B_{\mathrm{true}}=4000$. \textit{(continued)}}\\
\toprule
$k$ & $m$ & $n$ & $\rho$ & $S(t)$ & $\widehat S_{\mathrm{RSS}}$ & $\widehat S_{\mathrm{SRS}}$ & $\mathrm{RE}_{\mathrm{true}}$ & $\mathrm{RE}_{\mathrm{MC}}$ & $\mathrm{RE}_{\mathrm{GW}}$\\
\midrule
\endhead

\endfoot
\bottomrule
\endlastfoot
\cellcolor{gray!10}{2} & \cellcolor{gray!10}{20} & \cellcolor{gray!10}{40} & \cellcolor{gray!10}{0.1} & \cellcolor{gray!10}{0.10} & \cellcolor{gray!10}{0.132} & \cellcolor{gray!10}{0.108} & \cellcolor{gray!10}{1.262} & \cellcolor{gray!10}{1.268} & \cellcolor{gray!10}{0.913}\\
2 & 20 & 40 & 0.3 & 0.10 & 0.134 & 0.108 & 1.218 & 1.190 & 0.919\\
\cellcolor{gray!10}{2} & \cellcolor{gray!10}{20} & \cellcolor{gray!10}{40} & \cellcolor{gray!10}{0.5} & \cellcolor{gray!10}{0.10} & \cellcolor{gray!10}{0.138} & \cellcolor{gray!10}{0.107} & \cellcolor{gray!10}{1.097} & \cellcolor{gray!10}{1.067} & \cellcolor{gray!10}{0.892}\\
2 & 20 & 40 & 0.7 & 0.10 & 0.138 & 0.108 & 1.144 & 1.126 & 0.895\\
\cellcolor{gray!10}{2} & \cellcolor{gray!10}{20} & \cellcolor{gray!10}{40} & \cellcolor{gray!10}{0.9} & \cellcolor{gray!10}{0.10} & \cellcolor{gray!10}{0.139} & \cellcolor{gray!10}{0.108} & \cellcolor{gray!10}{1.170} & \cellcolor{gray!10}{1.087} & \cellcolor{gray!10}{0.903}\\
\addlinespace
2 & 50 & 100 & 0.1 & 0.10 & 0.104 & 0.100 & 1.136 & 1.182 & 1.005\\
\cellcolor{gray!10}{2} & \cellcolor{gray!10}{50} & \cellcolor{gray!10}{100} & \cellcolor{gray!10}{0.3} & \cellcolor{gray!10}{0.10} & \cellcolor{gray!10}{0.107} & \cellcolor{gray!10}{0.100} & \cellcolor{gray!10}{1.115} & \cellcolor{gray!10}{1.098} & \cellcolor{gray!10}{0.993}\\
2 & 50 & 100 & 0.5 & 0.10 & 0.113 & 0.100 & 1.059 & 1.097 & 0.982\\
\end{longtable}


\begin{longtable}[t]{rrrrrrrrrr}
\caption{\label{tab:tab:aft_pc50_s75}t=0.35, S(t)=75\% --- AFT superpopulation: results at 50\% censoring. Columns show KM estimates for RSS and SRS and three measures of relative efficiency (true, Monte Carlo, Greenwood). Replicates: $B_{\mathrm{MC}}=10000$, $B_{\mathrm{true}}=4000$.}\\
\toprule
$k$ & $m$ & $n$ & $\rho$ & $S(t)$ & $\widehat S_{\mathrm{RSS}}$ & $\widehat S_{\mathrm{SRS}}$ & $\mathrm{RE}_{\mathrm{true}}$ & $\mathrm{RE}_{\mathrm{MC}}$ & $\mathrm{RE}_{\mathrm{GW}}$\\
\midrule
\endfirsthead
\caption[]{t=0.35, S(t)=75\% --- AFT superpopulation: results at 50\% censoring. Columns show KM estimates for RSS and SRS and three measures of relative efficiency (true, Monte Carlo, Greenwood). Replicates: $B_{\mathrm{MC}}=10000$, $B_{\mathrm{true}}=4000$. \textit{(continued)}}\\
\toprule
$k$ & $m$ & $n$ & $\rho$ & $S(t)$ & $\widehat S_{\mathrm{RSS}}$ & $\widehat S_{\mathrm{SRS}}$ & $\mathrm{RE}_{\mathrm{true}}$ & $\mathrm{RE}_{\mathrm{MC}}$ & $\mathrm{RE}_{\mathrm{GW}}$\\
\midrule
\endhead

\endfoot
\bottomrule
\endlastfoot
\cellcolor{gray!10}{2} & \cellcolor{gray!10}{20} & \cellcolor{gray!10}{40} & \cellcolor{gray!10}{0.1} & \cellcolor{gray!10}{0.75} & \cellcolor{gray!10}{0.751} & \cellcolor{gray!10}{0.751} & \cellcolor{gray!10}{0.990} & \cellcolor{gray!10}{1.011} & \cellcolor{gray!10}{1.032}\\
2 & 20 & 40 & 0.3 & 0.75 & 0.748 & 0.748 & 1.084 & 1.051 & 1.089\\
\cellcolor{gray!10}{2} & \cellcolor{gray!10}{20} & \cellcolor{gray!10}{40} & \cellcolor{gray!10}{0.5} & \cellcolor{gray!10}{0.75} & \cellcolor{gray!10}{0.752} & \cellcolor{gray!10}{0.751} & \cellcolor{gray!10}{1.194} & \cellcolor{gray!10}{1.171} & \cellcolor{gray!10}{1.232}\\
2 & 20 & 40 & 0.7 & 0.75 & 0.751 & 0.749 & 1.191 & 1.184 & 1.241\\
\cellcolor{gray!10}{2} & \cellcolor{gray!10}{20} & \cellcolor{gray!10}{40} & \cellcolor{gray!10}{0.9} & \cellcolor{gray!10}{0.75} & \cellcolor{gray!10}{0.750} & \cellcolor{gray!10}{0.751} & \cellcolor{gray!10}{1.169} & \cellcolor{gray!10}{1.143} & \cellcolor{gray!10}{1.237}\\
\addlinespace
2 & 50 & 100 & 0.1 & 0.75 & 0.750 & 0.751 & 1.021 & 1.001 & 1.016\\
\cellcolor{gray!10}{2} & \cellcolor{gray!10}{50} & \cellcolor{gray!10}{100} & \cellcolor{gray!10}{0.3} & \cellcolor{gray!10}{0.75} & \cellcolor{gray!10}{0.750} & \cellcolor{gray!10}{0.751} & \cellcolor{gray!10}{1.060} & \cellcolor{gray!10}{1.055} & \cellcolor{gray!10}{1.068}\\
2 & 50 & 100 & 0.5 & 0.75 & 0.752 & 0.751 & 1.199 & 1.177 & 1.221\\
\end{longtable}


\begin{longtable}[t]{rrrrrrrrrr}
\caption{\label{tab:tab:aft_pc50_s50}t=1.00, S(t)=50\% --- AFT superpopulation: results at 50\% censoring. Columns show KM estimates for RSS and SRS and three measures of relative efficiency (true, Monte Carlo, Greenwood). Replicates: $B_{\mathrm{MC}}=10000$, $B_{\mathrm{true}}=4000$.}\\
\toprule
$k$ & $m$ & $n$ & $\rho$ & $S(t)$ & $\widehat S_{\mathrm{RSS}}$ & $\widehat S_{\mathrm{SRS}}$ & $\mathrm{RE}_{\mathrm{true}}$ & $\mathrm{RE}_{\mathrm{MC}}$ & $\mathrm{RE}_{\mathrm{GW}}$\\
\midrule
\endfirsthead
\caption[]{t=1.00, S(t)=50\% --- AFT superpopulation: results at 50\% censoring. Columns show KM estimates for RSS and SRS and three measures of relative efficiency (true, Monte Carlo, Greenwood). Replicates: $B_{\mathrm{MC}}=10000$, $B_{\mathrm{true}}=4000$. \textit{(continued)}}\\
\toprule
$k$ & $m$ & $n$ & $\rho$ & $S(t)$ & $\widehat S_{\mathrm{RSS}}$ & $\widehat S_{\mathrm{SRS}}$ & $\mathrm{RE}_{\mathrm{true}}$ & $\mathrm{RE}_{\mathrm{MC}}$ & $\mathrm{RE}_{\mathrm{GW}}$\\
\midrule
\endhead

\endfoot
\bottomrule
\endlastfoot
\cellcolor{gray!10}{2} & \cellcolor{gray!10}{20} & \cellcolor{gray!10}{40} & \cellcolor{gray!10}{0.1} & \cellcolor{gray!10}{0.50} & \cellcolor{gray!10}{0.499} & \cellcolor{gray!10}{0.498} & \cellcolor{gray!10}{0.930} & \cellcolor{gray!10}{1.045} & \cellcolor{gray!10}{1.034}\\
2 & 20 & 40 & 0.3 & 0.50 & 0.500 & 0.500 & 1.034 & 1.086 & 1.100\\
\cellcolor{gray!10}{2} & \cellcolor{gray!10}{20} & \cellcolor{gray!10}{40} & \cellcolor{gray!10}{0.5} & \cellcolor{gray!10}{0.50} & \cellcolor{gray!10}{0.501} & \cellcolor{gray!10}{0.500} & \cellcolor{gray!10}{1.318} & \cellcolor{gray!10}{1.301} & \cellcolor{gray!10}{1.300}\\
2 & 20 & 40 & 0.7 & 0.50 & 0.502 & 0.500 & 1.306 & 1.231 & 1.310\\
\cellcolor{gray!10}{2} & \cellcolor{gray!10}{20} & \cellcolor{gray!10}{40} & \cellcolor{gray!10}{0.9} & \cellcolor{gray!10}{0.50} & \cellcolor{gray!10}{0.499} & \cellcolor{gray!10}{0.500} & \cellcolor{gray!10}{1.245} & \cellcolor{gray!10}{1.297} & \cellcolor{gray!10}{1.314}\\
\addlinespace
2 & 50 & 100 & 0.1 & 0.50 & 0.499 & 0.500 & 1.028 & 0.998 & 1.018\\
\cellcolor{gray!10}{2} & \cellcolor{gray!10}{50} & \cellcolor{gray!10}{100} & \cellcolor{gray!10}{0.3} & \cellcolor{gray!10}{0.50} & \cellcolor{gray!10}{0.500} & \cellcolor{gray!10}{0.502} & \cellcolor{gray!10}{1.069} & \cellcolor{gray!10}{1.051} & \cellcolor{gray!10}{1.083}\\
2 & 50 & 100 & 0.5 & 0.50 & 0.500 & 0.500 & 1.288 & 1.255 & 1.293\\
\end{longtable}


\begin{longtable}[t]{rrrrrrrrrr}
\caption{\label{tab:tab:aft_pc50_s25}t=2.85, S(t)=25\% --- AFT superpopulation: results at 50\% censoring. Columns show KM estimates for RSS and SRS and three measures of relative efficiency (true, Monte Carlo, Greenwood). Replicates: $B_{\mathrm{MC}}=10000$, $B_{\mathrm{true}}=4000$.}\\
\toprule
$k$ & $m$ & $n$ & $\rho$ & $S(t)$ & $\widehat S_{\mathrm{RSS}}$ & $\widehat S_{\mathrm{SRS}}$ & $\mathrm{RE}_{\mathrm{true}}$ & $\mathrm{RE}_{\mathrm{MC}}$ & $\mathrm{RE}_{\mathrm{GW}}$\\
\midrule
\endfirsthead
\caption[]{t=2.85, S(t)=25\% --- AFT superpopulation: results at 50\% censoring. Columns show KM estimates for RSS and SRS and three measures of relative efficiency (true, Monte Carlo, Greenwood). Replicates: $B_{\mathrm{MC}}=10000$, $B_{\mathrm{true}}=4000$. \textit{(continued)}}\\
\toprule
$k$ & $m$ & $n$ & $\rho$ & $S(t)$ & $\widehat S_{\mathrm{RSS}}$ & $\widehat S_{\mathrm{SRS}}$ & $\mathrm{RE}_{\mathrm{true}}$ & $\mathrm{RE}_{\mathrm{MC}}$ & $\mathrm{RE}_{\mathrm{GW}}$\\
\midrule
\endhead

\endfoot
\bottomrule
\endlastfoot
\cellcolor{gray!10}{2} & \cellcolor{gray!10}{20} & \cellcolor{gray!10}{40} & \cellcolor{gray!10}{0.1} & \cellcolor{gray!10}{0.25} & \cellcolor{gray!10}{0.255} & \cellcolor{gray!10}{0.250} & \cellcolor{gray!10}{1.033} & \cellcolor{gray!10}{1.124} & \cellcolor{gray!10}{1.030}\\
2 & 20 & 40 & 0.3 & 0.25 & 0.258 & 0.252 & 1.166 & 1.184 & 1.065\\
\cellcolor{gray!10}{2} & \cellcolor{gray!10}{20} & \cellcolor{gray!10}{40} & \cellcolor{gray!10}{0.5} & \cellcolor{gray!10}{0.25} & \cellcolor{gray!10}{0.266} & \cellcolor{gray!10}{0.249} & \cellcolor{gray!10}{1.281} & \cellcolor{gray!10}{1.295} & \cellcolor{gray!10}{1.120}\\
2 & 20 & 40 & 0.7 & 0.25 & 0.269 & 0.251 & 1.344 & 1.270 & 1.130\\
\cellcolor{gray!10}{2} & \cellcolor{gray!10}{20} & \cellcolor{gray!10}{40} & \cellcolor{gray!10}{0.9} & \cellcolor{gray!10}{0.25} & \cellcolor{gray!10}{0.267} & \cellcolor{gray!10}{0.250} & \cellcolor{gray!10}{1.248} & \cellcolor{gray!10}{1.272} & \cellcolor{gray!10}{1.132}\\
\addlinespace
2 & 50 & 100 & 0.1 & 0.25 & 0.249 & 0.249 & 1.062 & 0.984 & 1.019\\
\cellcolor{gray!10}{2} & \cellcolor{gray!10}{50} & \cellcolor{gray!10}{100} & \cellcolor{gray!10}{0.3} & \cellcolor{gray!10}{0.25} & \cellcolor{gray!10}{0.252} & \cellcolor{gray!10}{0.251} & \cellcolor{gray!10}{1.065} & \cellcolor{gray!10}{1.056} & \cellcolor{gray!10}{1.065}\\
2 & 50 & 100 & 0.5 & 0.25 & 0.254 & 0.251 & 1.179 & 1.254 & 1.175\\
\end{longtable}


\begin{longtable}[t]{rrrrrrrrrr}
\caption{\label{tab:tab:aft_pc50_s10}t=7.30, S(t)=10\% --- AFT superpopulation: results at 50\% censoring. Columns show KM estimates for RSS and SRS and three measures of relative efficiency (true, Monte Carlo, Greenwood). Replicates: $B_{\mathrm{MC}}=10000$, $B_{\mathrm{true}}=4000$.}\\
\toprule
$k$ & $m$ & $n$ & $\rho$ & $S(t)$ & $\widehat S_{\mathrm{RSS}}$ & $\widehat S_{\mathrm{SRS}}$ & $\mathrm{RE}_{\mathrm{true}}$ & $\mathrm{RE}_{\mathrm{MC}}$ & $\mathrm{RE}_{\mathrm{GW}}$\\
\midrule
\endfirsthead
\caption[]{t=7.30, S(t)=10\% --- AFT superpopulation: results at 50\% censoring. Columns show KM estimates for RSS and SRS and three measures of relative efficiency (true, Monte Carlo, Greenwood). Replicates: $B_{\mathrm{MC}}=10000$, $B_{\mathrm{true}}=4000$. \textit{(continued)}}\\
\toprule
$k$ & $m$ & $n$ & $\rho$ & $S(t)$ & $\widehat S_{\mathrm{RSS}}$ & $\widehat S_{\mathrm{SRS}}$ & $\mathrm{RE}_{\mathrm{true}}$ & $\mathrm{RE}_{\mathrm{MC}}$ & $\mathrm{RE}_{\mathrm{GW}}$\\
\midrule
\endhead

\endfoot
\bottomrule
\endlastfoot
\cellcolor{gray!10}{2} & \cellcolor{gray!10}{20} & \cellcolor{gray!10}{40} & \cellcolor{gray!10}{0.1} & \cellcolor{gray!10}{0.10} & \cellcolor{gray!10}{0.167} & \cellcolor{gray!10}{0.126} & \cellcolor{gray!10}{1.176} & \cellcolor{gray!10}{1.148} & \cellcolor{gray!10}{0.958}\\
2 & 20 & 40 & 0.3 & 0.10 & 0.166 & 0.126 & 1.109 & 1.069 & 0.961\\
\cellcolor{gray!10}{2} & \cellcolor{gray!10}{20} & \cellcolor{gray!10}{40} & \cellcolor{gray!10}{0.5} & \cellcolor{gray!10}{0.10} & \cellcolor{gray!10}{0.168} & \cellcolor{gray!10}{0.128} & \cellcolor{gray!10}{1.051} & \cellcolor{gray!10}{1.042} & \cellcolor{gray!10}{0.940}\\
2 & 20 & 40 & 0.7 & 0.10 & 0.167 & 0.126 & 1.102 & 1.050 & 0.940\\
\cellcolor{gray!10}{2} & \cellcolor{gray!10}{20} & \cellcolor{gray!10}{40} & \cellcolor{gray!10}{0.9} & \cellcolor{gray!10}{0.10} & \cellcolor{gray!10}{0.165} & \cellcolor{gray!10}{0.128} & \cellcolor{gray!10}{1.018} & \cellcolor{gray!10}{1.084} & \cellcolor{gray!10}{0.950}\\
\addlinespace
2 & 50 & 100 & 0.1 & 0.10 & 0.119 & 0.104 & 1.299 & 1.202 & 1.019\\
\cellcolor{gray!10}{2} & \cellcolor{gray!10}{50} & \cellcolor{gray!10}{100} & \cellcolor{gray!10}{0.3} & \cellcolor{gray!10}{0.10} & \cellcolor{gray!10}{0.121} & \cellcolor{gray!10}{0.105} & \cellcolor{gray!10}{1.111} & \cellcolor{gray!10}{1.169} & \cellcolor{gray!10}{1.016}\\
2 & 50 & 100 & 0.5 & 0.10 & 0.123 & 0.104 & 1.042 & 1.072 & 0.982\\
\end{longtable}

\end{document}